\documentclass{article}

\makeatletter
\usepackage{yeyt}

\usepackage{hyperref}
\usepackage{authblk}
\usepackage[noend]{algpseudocode}
\usepackage[dvipsnames]{xcolor}
\usepackage{listings,enumerate,color,relsize,multirow}
\usepackage[counterclockwise]{rotating}
\usepackage[utf8]{inputenc}
\usepackage{pifont,xr}
\usepackage[symbol]{footmisc}

\newcommand{\beq}{\begin{equation}}
\newcommand{\eeq}{\end{equation}}
\newcommand{\beas}{\begin{eqnarray*}}
\newcommand{\eeas}{\end{eqnarray*}}
\newcommand{\bea}{\begin{eqnarray}}
\newcommand{\eea}{\end{eqnarray}}
\newcommand{\cmark}{\ding{51}}
\newcommand{\xmark}{\ding{55}}

\newcommand{\vect}{\mathrm{vec}}
\newbox\TempBox \newbox\TempBoxA
\newcommand{\non}{\nonumber \\}
\def\Cov{{\mathrm{cov}}} 
\def\Var{{\mathrm{var}}} 

\def\T{{ \mathrm{\scriptscriptstyle T} }}

\newcommand{\real}[1]{\mathrm{I \! R} \mathit{^{#1}}}

\def\A{\boldsymbol{A}}

\def\D{\boldsymbol{D}}

\def\R{\boldsymbol{R}}

\def\X{\boldsymbol{X}}
\def\Y{\boldsymbol{Y}}
\def\Z{\boldsymbol{Z}}

\def\L12{L_{12}}
\def\P{\mathbb{P}}
\newcommand{\Rho}{\boldsymbol{P}}

\def\be{\boldsymbol{\beta}}
\def\hbe{\hat{\boldsymbol{\beta}}}

\def\S{\boldsymbol{\Sigma}}
\def\hSigma{\hat{\boldsymbol{\Sigma}}}
\def\O{\boldsymbol{\Omega}}
\def\m{\boldsymbol{\mu}}
\def\Bnu{\boldsymbol{\nu}}
\def\eps{\boldsymbol{\epsilon}}
\def\teps{\tilde{\epsilon}}
\def\beps{\bar{\epsilon}}
\def\heps{\hat{\epsilon}}

\def\hsigma{\hat{\sigma}}

\begin{document}
\title{Paired Test of Matrix Graphs and Brain Connectivity Analysis}

\author{YUTING YE$^{\dagger}$ \\
\textit{Division of Biostatistics, University of California at Berkeley, Berkeley, CA 94720, USA} \\
YIN XIA$^{\dagger}$ \\
\textit{Department of Statistics, School of Management, Fudan University, Shanghai, 200433, China}\\
LEXIN LI$^*$ \\ 
\textit{Division of Biostatistics, University of California at Berkeley, Berkeley, CA 94720, USA} \\
lexinli@berkeley.edu\\
}


\markboth%
{Y. Ye and others}
{Paired Test of Matrix Graphs}

\footnotetext[2]{The first two authors share the co-first authorship.}
\footnotetext{To whom correspondence should be addressed.}
\date{\today} 

\maketitle

\begin{abstract}
{Inferring brain connectivity network and quantifying the significance of interactions between brain regions are of paramount importance in neuroscience. Although there have recently emerged some tests for graph inference based on independent samples, there is no readily available solution to test the change of brain network for paired and correlated samples. In this article, we develop a paired test of matrix graphs to infer brain connectivity network when the groups of samples are correlated. The proposed test statistic is both bias corrected and variance corrected, and achieves a small estimation error rate. The subsequent multiple testing procedure built on this test statistic is guaranteed to asymptotically control the false discovery rate at the pre-specified level. Both the methodology and theory of the new test are considerably different from the two independent samples framework, owing to the strong correlations of measurements on the same subjects before and after the stimulus activity. We illustrate the efficacy of our proposal through simulations and an analysis of an Alzheimer's Disease Neuroimaing Initiative dataset.}
\end{abstract}

{\textbf{Keywords:} Brain connectivity analysis; Gaussian graphical model; Matrix-variate normal distribution; Multiple testing; Partial correlation; Variance correction.}


\section{Introduction}
\label{sec:intro}

Brain functional connectivity reveals the intrinsic functional architecture of brains by measuring correlations in neurophysiological recordings of brain activities (\citealp{Varoquaux2013}). Numerous studies have found that functional connectivity alters for individuals with neurological disorders, such as Alzheimer's diseases (AD) and autism spectrum disorder (\citealp{hedden2009disruption, rudie2013altered}), or after experiencing stimulus activities such as stress or therapy (\citealp{peck2004functional, Marle2010}). The brain connectivity network is believed to hold crucial insight to help understand the pathologies of neurological disorders and to develop targeting treatment (\citealp{fox2010clinical, Quaedflieg2015}).  

Brain functional connectivity is commonly encoded as a network, or graph, with nodes representing brain regions, and links representing interactions and correlations between regions. Among multiple correlation measures, partial correlation is a well accepted and frequently used metric, and correspondingly, the connectivity network is portrayed by a partial correlation matrix (\citealp{ryali2012estimation, chen2013estimation}). Current mainstream imaging modalities to study functional connectivity include electroencephalography (EEG),  electrocorticography (ECoG), and resting-state functional magnetic resonance imaging (fMRI). After proper preprocessing, the resulting imaging data for each subject is summarized in the form of a location by time matrix, from which a partial correlation matrix is constructed to characterize brain connectivity. 

A central problem in connectivity analysis is inference. Unlike network \emph{estimation} (\citealp{ahn2015sparse, chen2015parsimonious, kang2016depression, qiu2016joint, wang2016efficient}), network \emph{inference} aims to directly quantify the statistical significance of individual links or their differences, meanwhile explicitly controlling for the false discovery. Recently there have been proposals of partial correlation matrix based network inference for vector-valued data following a normal distribution (\citealp{liu2013gaussian, xia2015testing}), or matrix-valued data following a matrix-normal distribution (\citealp{ChenLiu2015, xia2017one, xia2018two}). For brain connectivity analysis, the data obtained from EEG, ECoG, or fMRI is of a matrix form, and the primary scientific interest is on the spatial but not the temporal correlation patterns of the brain. Directly applying the tests for vector-valued data to infer the spatial patterns ignores the temporal correlations among the columns of the matrix data, and is to result in distorted test size and false discovery rate (\citealp{xia2017one}). Whitening can alleviate this problem, and in effect transforms the matrix data back to the vector case (\citealp{narayan2015two}). However, it does not utilize the data efficiently, would result in loss of power, and is also computationally intensive (\citealp{xia2018two}). Alternatively, \citet{ChenLiu2015, xia2017one} directly tackled inference of the matrix-valued data under the one-sample testing scenario, and \citet{xia2018two} tackled the two-sample scenario where the two groups of samples are independent. 

In addition to inference about brain network alternation across independent subject groups, it is of equal interest and importance to infer the change of brain network of the same group of subjects before and after a ``stimulus" activity, which could be a treatment, a disease conversion, or a different experimental condition.  For instance, \citet{peck2004functional} studied brain connectivity activities in auditory and motor cortices of aphasic patients before and after a therapy. \citet{Gianaros2008, Marle2010, Quaedflieg2015} studied amygdala-centered connectivity patterns in healthy subjects before and after the experimentally-induced stress. \citet{cai2015network} studied alterations in brain functional networks in patients with primary angle-closure glaucoma before and after the surgery. \citet{kang2016decrease} studied brain connectivity activities in left and right inferior frontal gyrus areas of the same subjects under different sleeping conditions. \citet{Ficek2018} studied changes of functional connectivity before and after a language intervention therapy. In Section \ref{sec:realdata}, we aim to identify the connectivity patterns that differ before and after a patient converted to AD. The two-sample test of \citet{xia2018two} does not directly apply to those studies, because of the strong correlations of brain measurements on the \emph{same} subjects before and after the stimulus. For instance, a positive correlation before and after the stimulus would reduce the variance of the partial correlation difference between the two groups, causing the two-sample test to overestimate the variance and resulting in a low test power. On the contrary, a negative correlation would inflate the variance, causing the two-sample test to underestimate the variance and resulting in an inflated false discovery. 

In this article, we develop a paired test of matrix graphs to infer brain connectivity network when the groups of samples are correlated, such as in the scenario of before and after the stimulus. The key of our proposal is an innovative variance correction procedure that incorporates the spatial and temporal dependency between the paired samples. The proposed test statistic is both bias corrected and variance corrected, and is shown to achieve a sufficiently small estimation error rate asymptotically. This in turn ensures that the subsequent multiple testing procedure built on this test statistic can asymptotically control the false discovery rate at the pre-specified level. Our proposal extends the two-sample test of \citet{xia2018two}, but is considerably different. This extension is far from trivial, and the theoretical investigation of the paired test is much more involved, as one needs to carefully evaluate both within-sample and between-sample correlations. To our knowledge, there is no existing graph inference procedure for paired matrix samples, and our proposal offers a timely response to an important problem of both scientific and methodological interest.   

The rest of the article is organized as follows. Section \ref{sec:method} presents the formulation of the hypothesis testing problem, the proposed test statistic, the variance correction procedure for the paired samples, and the multiple testing procedure. Section \ref{sec:theory:results} studies the corresponding asymptotic properties. Section \ref{sec:simulation} examines the empirical performance of the proposed test through simulations, and Section \ref{sec:realdata} analyzes a real fMRI dataset. The appendix collects all the technical assumptions, proofs and additional numerical results.

\section{Paired test}
\label{sec:method}

\subsection{Problem formulation}

Let $\X^{(t)}$ denote the $p \times q$ matrix observed at time point $t, t=1,2$. In brain connectivity analysis, $\X^{(t)}$ denotes the spatial-temporal imaging data before ($t=1$) and after ($t=2$) a stimulus activity or conversion, and each $\X^{(t)}$ corresponds to $p$ brain regions and the time course data of each region is of length $q$. We assume $\{ \X^{(1)}, \X^{(2)} \}$ follows a matrix normal distribution, i.e., 
\begin{equation} \label{eq:joint_dist_X}
\left(
\begin{array}{c}
\vect \{ \X^{(1)} \}  \\
\vect \{ \X^{(2)} \} 
\end{array}
\right) \sim
\mathrm{Normal}
\left( \bold{0}_{2pq},
\S
\right), \mbox{ with } 
\S=\left(
\begin{array}{cc}
\S_{S_1} \otimes \S_{T_1} &     \S_{S_{1,2}} \otimes \S_{T_{1,2}}\\
\S_{S_{1,2}}^{\T} \otimes \S_{T_{1,2}}^{\T} & \S_{S_2} \otimes \S_{T_2}
\end{array}
\right).
\smallskip
\end{equation}
Without loss of generality, the mean is assumed to be zero, $\otimes$ is the Kronecker product, and $\vect(\cdot)$ is the operator that stacks the columns of a matrix into a vector. Furthermore, $\S_{S_t} \in \real{p \times p}$ denotes the covariance matrix of the spatial regions, $\S_{T_t} \in \real{q \times q}$ denotes the temporal covariance matrix of the time course data, at $t=1,2$, respectively, and $\S_{S_{1,2}}$ and $\S_{T_{1,2}}$ denote the between-sample spatial and temporal covariance, respectively. When $\S_{S_{1,2}} \otimes \S_{T_{1,2}} = \bold 0$, \eqref{eq:joint_dist_X} reduces to the independent two-sample setting of \citet{xia2018two}. We remark that, the matrix normal distribution has been frequently adopted in numerous applications involving matrix-valued data (\citealp{YinLi2012, LengTang2012}), and is also  scientifically plausible in neuroimaging analysis (\citealp{Smith2004, Friston2007}). Moreover, \cite{aston2017tests} developed a test to check if the
data conforms with the Kronecker product structure. In Section \ref{sec:sensitivity}, we further carry out sensitivity analysis, and show that our proposed test works reasonably well even when the data deviates from the matrix normal distribution \eqref{eq:joint_dist_X}.
 
Let $\O_{S_t} = \S_{S_t}^{-1} = (\omega_{S_t,i,j})_{i,j=1}^{p}$ denote the spatial precision matrix, $\D_{S_t}$ denote the diagonal matrix of $\O_{S_t}$ and $\R_{S_t}=\D_{S_t}^{-1/2}\O_{S_t}\D_{S_t}^{-1/2} = (\rho_{S_t,i,j})_{i,j=1}^{p}$ denote the spatial partial correlation matrix. In brain connectivity analysis, the primary interest is to infer the connectivity network characterized by the spatial partial correlation matrix. The temporal covariance or precision matrix is of little interest in this context and is to be treated as a nuisance parameter. Consequently, we formulate our inference problem as simultaneously testing 
\begin{eqnarray}\label{eq:entry.test}
H_{0,i,j}: \rho_{S_{1},i,j}= \rho_{S_{2},i,j} \;\; \mathrm{ versus } \;\; H_{1,i,j}: \rho_{S_{1},i,j} \neq \rho_{S_{2},i,j}, \quad \textrm{ for } \; 1 \leq i < j \leq p.
\end{eqnarray}
We next derive the test statistic and the associated variance correction to account for the correlations of the paired samples.

\subsection{Test statistic}
\label{sec:method:test_stat}

Consider $n$ pairs of samples $\left\{ \X^{(1)}_k, \X^{(2)}_k \right\}_{k=1}^{n}$ from the joint distribution \eqref{eq:joint_dist_X}. To construct the test statistic for \eqref{eq:entry.test}, we first remove the temporal correlations by the linear transformation $\Y^{(t)}_k = \X^{(t)}_{k}{\S}_{T_t}^{-1/2}$, $k=1,\ldots,n$, $t=1, 2$, and  
\begin{equation} \label{eq:joint_dist_Y}
\smallskip
\left (
\begin{array}{c}
\vect \{ \Y^{(1)} \} \\
\vect \{ \Y^{(2)} \} 
\end{array}
\right ) \sim
\mathrm{Normal}
\left( \bold{0}_{2pq},
\left(
\begin{array}{cc}
\S_{S_1} \otimes \boldsymbol{I}_q &     \S_{S_{1,2}} \otimes \Rho_{T_{1,2}}\\
\S_{S_{1,2}}^{\T} \otimes \Rho_{T_{1,2}}^{\T} & \S_{S_2} \otimes \boldsymbol{I}_q
\end{array}
\right )
\right),
\smallskip
\end{equation}
where $\Rho_{T_{1,2}} = \S_{T_1}^{-1/2} \S_{T_{1,2}} \S_{T_2}^{-1/2}$ denotes the between-sample temporal covariance matrix of the transformed samples. Clearly, for the independent case, $\S_{S_{1,2}} \otimes \Rho_{T_{1,2}} = 0$. In practice, $\S_{T_t}$ and $\S_{T_{1,2}}$ are generally unknown. There are multiple ways to estimate $\S_{T_t}$, or equivalently, $\O_{T_t} = \S_{T_t}^{-1}$. Examples include the sample covariance estimator, the banded covariance estimator (\citealp{bickel2008regularized}), the adaptive thresholding estimator (\citealp{cai2011adaptive}) for $\S_{T_t}$, or the Clime estimator (\citealp{cai2011constrained}) for $\O_{T_t}$. We adopt the banded estimator in this article, given its competitive performance in both the one-sample test and the independent two-sample test under the matrix normal distribution (\citealp{xia2017one, xia2018two}). In the following, we first derive the test statistic with known $\S_{T_t}$ and $\Rho_{T_{1,2}}$, which helps simplify the notations considerably. We then extend it by plugging in an estimated $\S_{T_t}$ and $\Rho_{T_{1,2}}$. Accordingly, we will add the superscript $(d)$ in the resulting statistics to represent this scenario when $\S_{T_t}$ and $\Rho_{T_{1,2}}$ are estimated given the data. In Section \ref{sec:theory:results} we show that the test statistics under the known $\S_{T_t}$, $\Rho_{T_{1,2}}$ and the estimated $\S_{T_t}$, $\Rho_{T_{1,2}}$ have the same asymptotic property. Consequently, they lead to the same multiple testing procedure with the guaranteed asymptotic control of false discovery. 

The construction of our test statistic is based on the fact that, under the normal distribution, the precision matrix can be described through the regression model (\citealp{anderson2003}),
\begin{eqnarray} \label{eq:inv_reg}
Y^{(t)}_{k,i,l} = \Y_{k,-i,l}^{{(t)}\T}\be_{i}^{(t)}+\epsilon^{(t)}_{k,i,l}, \quad 1\leq i \leq p, ~ 1\leq l \leq q, ~ 1 \leq k \leq n, ~ t = 1, 2,
\end{eqnarray}
where the error term $\epsilon^{(t)}_{k,i,l} \sim N\left(0,\sigma_{S_t,i,i}-\S_{S_t,i,-i}\S_{S_t,-i,-i}^{-1}\S_{S_t,-i,i}\right)$ and is independent of $\Y^{(t)}_{k,-i,l}$, and the subscript $-i$ means the $i$th entry is removed from a vector, or the $i$th row or column removed from a matrix. The regression coefficient $\be_{i}^{(t)}$ can be estimated using Lasso or other methods, as long as the estimator $\hat\be_{i}^{(t)}$ satisfies the regularity condition (A5) or (A6) in the appendix. See \citet{xia2017one, xia2018two} and Section \ref{appendix:tune_Lasso} of the appendix for a more detailed discussion on estimation of $\be_{i}^{(t)}$ and the associated tuning procedure. Moreover, the error term satisfies that $r_{i,j}^{(t)} = \Cov \left\{ \epsilon^{(t)}_{k,i,l},\epsilon^{(t)}_{k,j,l} \right\}= \omega_{S_t,i,j} / (\omega_{S_t,i,i}\omega_{S_t,j,j})$. Therefore the element $\omega_{S_t,i,j}$ of the spatial precision matrix $\R_{S_t}$, and in turn, the element $\rho_{S_t,i,j}$ of the spatial partial correlation matrix $\R_{S_t}$ can be represented in terms of $r_{i,j}^{(t)}$. Following \citet{xia2017one, xia2018two}, a bias-corrected estimator of $r_{i,j}^{(t)}$ is obtained from fitting the regression model \eqref{eq:inv_reg}, 
\begin{eqnarray*}
\hat{r}_{i,j}^{(t)} = 
\begin{cases}
- \tilde{r}_{i,j}^{(t)} - \tilde{r}_{i,i}^{(t)}\hat{\beta}_{i,j}^{(t)} - \tilde{r}_{j,j}^{(t)}\hat{\beta}_{j-1,i}^{(t)}, & \mathrm{when} \;\; 1 \leq i < j \leq p \\
\tilde{r}_{i,i}^{(t)},  & \mathrm{when}\;\; 1 \leq i = j \leq p,
\end{cases}
\end{eqnarray*}
where $\tilde r_{i,j}^{(t)} = (nq)^{-1} \sum_{k=1}^{n}\sum_{l=1}^q\hat{\epsilon}^{(t)}_{k,i,l}\hat{\epsilon}^{(t)}_{k,j,l}$ is the sample covariance between the residuals, $\hat{\epsilon}^{(t)}_{k,i,l}=Y^{(t)}_{k,i,l}-\bar{Y}^{(t)}_{i,l}-(\Y^{(t)}_{k,-i,l}-\bar{\Y}^{(t)}_{\cdot,-i,l})^{\T}\hat{\be}_{i}^{(t)}$, $\bar{Y}_{i,l}^{(t)}=n^{-1} \sum_{k=1}^n Y_{k,i,l}^{(t)}$, and $\bar{\Y}^{(t)}_{\cdot,-i,l}=n^{-1} \sum_{k=1}^n\Y^{(t)}_{k,-i,l}$. Based on the estimator $\hat{r}_{i,j}^{(t)}$, we further obtain a bias-corrected estimator of the element $\rho_{S_t,i,j}$ of the spatial partial correlation matrix $\R_{S_t}$ as 
\begin{eqnarray*}
\hat \rho_{S_t,i,j} = \hat{r}_{i,j}^{(t)}/\{ \hat{r}_{i,i}^{(t)} \hat{r}_{j,j}^{(t)} \}^{1/2}, \quad 1\leq i < j\leq p, ~ t=1,2.
\end{eqnarray*}
We then construct our test statistic for the pair of hypotheses \eqref{eq:entry.test} as 
\begin{eqnarray*}\label{eq:test_W}
W_{i, j} = \frac{ (\hat \rho_{S_1,i,j} - \hat \rho_{S_2,i,j}) }{\hat \Theta_{i,j}^{1/2}}, \quad 1\leq i < j\leq p,
\end{eqnarray*}
where $\hat \Theta_{i,j}$ is an estimator of $\Var\left( \hat \rho_{S_1,i,j} - \hat \rho_{S_2,i,j} \right)$. We next develop such an estimator that incorporates the between-sample dependency of the paired samples.

\subsection{Variance correction} 
\label{sec:methods:var_correct}

We first recognize that the expression for the variance term $\Var\left( \hat \rho_{S_1,i,j} - \hat \rho_{S_2,i,j} \right)$ is quite involved. To alleviate this issue, we introduce an intermediate term, $\tilde{U}_{i,j}^{(t)} = \left\{ r_{i,j}^{(t)} - U_{i,j}^{(t)} \right\} / \left\{ r_{i,i}^{(t)}r_{j,j}^{(t)} \right\}^{\frac{1}{2}}$, where $U_{i,j}^{(t)} = (nq)^{-1} \sum_{k = 1}^{n}\sum_{l=1}^q\left[ \epsilon_{k,i,l}^{(t)}\epsilon_{k,j,l}^{(t)} - E\left\{ \epsilon_{k,i,l}^{(t)}\epsilon_{k,j,l}^{(t)} \right\} \right]$. Lemma \ref{lemma:hatrij_oracle} in the appendix implies that the difference between $\hat \rho_{S_t,i,j}$ and $\tilde{U}_{i,j}^{(t)}$ is negligible. Consequently, we estimate $\Var\left( \hat \rho_{S_1,i,j} - \hat \rho_{S_2,i,j} \right)$ by developing an estimator for 
\begin{eqnarray*}
\Theta_{i,j} = \Var\left\{ \tilde{U}_{i,j}^{(1)} - \tilde{U}_{i,j}^{(2)} \right\}.   
\end{eqnarray*}

For the independent two-sample setting, $\Theta_{i,j} = \theta_{i,j}^{(1)} + \theta_{i,j}^{(2)}$, where $\theta_{i,j}^{(t)} = \Var\left\{ \tilde{U}_{i,j}^{(t)} \right\}$, $t=1,2$. Based further on the observation that $\Var\left\{ \tilde{U}_{i,j}^{(t)} \right\} = \Var\left[ \epsilon^{(t)}_{k,i,l}\epsilon^{(t)}_{k,j,l} / \{ r_{i,i}^{(t)} r_{j,j}^{(t)} \}^{1/2} \right] / (nq) = \left[ 1+\{ \beta_{i,j}^{(t)} \}^2 r_{i,i}^{(t)} /  r_{j,j}^{(t)} \right] / (nq)$, we estimate $\theta_{i,j}^{(t)}$ by 
\begin{eqnarray} \label{theij}
\hat{\theta}_{i,j}^{(t)} = \frac{1}{nq} \left[ 1+ \left\{\hat{\beta}_{i,j}^{(t)}\right\}^2 \hat{r}_{i,i}^{(t)} / \hat{r}_{j,j}^{(t)} \right], \quad 1\leq i < j\leq p, ~ t=1,2.
\end{eqnarray}

For the paired samples, however, it is crucial to account for the between-sample spatial-temporal dependency as presented in $\S_{S_{1,2}}$ and $\S_{T_{1,2}}$ when estimating $\Theta_{i,j}$. Next we derive such an estimator of $\Theta_{i,j}$. Later in Section \ref{sec:theory:results}, we show that this estimator is accurate, in the sense that its scaled version achieves an $o_p(1/ \log p)$ convergence rate. This error rate is essential for the subsequent asymptotic false discovery control in multiple testing. 

The next proposition gives an explicit expression of $\Theta_{i,j}$ under the dependent setting. Its proof is given in the appendix. The key is the separable spatial and temporal dependence structures between the paired samples, and the decoupling of $\rho_{i,j;l_1, l_2}^{(1,2)}= \mathrm{corr}  \left\{ \epsilon_{k,i,l_1}^{(1)},\epsilon_{k,j,l_2}^{(2)} \right\}$ as $\rho_{i,j;l_1, l_2}^{(1,2)} = \rho_{S_{1,2,i,j}} \Rho_{T_{1,2,l_1,l_2}}$, where $\rho_{S_{1,2,i,j}}= \sqrt{r_{i,i}^{(1)}r_{j,j}^{(2)}} \; \O_{S_1, i, \cdot} \S_{S_{1,2}}\O_{S_2, \cdot, j}$ accounts for the spatial correlation, and $\Rho_{T_{1,2,l_1,l_2}}$ captures the temporal dependency. Here $\O_{S_1, i, \cdot}$  denotes the $i$th row of the matrix $\O_{S_1}$, and $\O_{S_2, \cdot, j}$ denotes the $j$th column of $\O_{S_2}$.
 
\begin{proposition}\label{thm:var_correct}
Under the data distribution (\ref{eq:joint_dist_Y}), we have, 
\begin{eqnarray}\label{eq:general_Theta}
\Theta_{i,j} &=& \theta_{i,j}^{(1)} + \theta_{i,j}^{(2)} - \frac{2}{nq^2} \bigl(\rho_{S_{1,2}, i, i}\rho_{S_{1,2}, j, j} + \rho_{S_{1,2}, i, j}\rho_{S_{1,2}, j, i} \bigr) \|\Rho_{T_{1,2}}\|_F^2 ,  
\end{eqnarray}
for $1 \leq i < j \leq p$, where $\| \cdot \|_F$ denotes the Frobenius norm. 
\end{proposition}

Define $\varrho_{i,j}^{(1,2)} = \rho_{S_{1,2}, i,j} \cdot \text{tr}(\Rho_{T_{1,2}})/q$, which is the correlation coefficient $\rho_{S_{1,2}, i,j}$ scaled by the term $\text{tr}(\Rho_{T_{1,2}})/q$, and $\text{tr}(\cdot)$ denotes the matrix trace. We observe that
\begin{eqnarray*}
\EE \left \{\frac{1}{nq} \sum_{k = 1}^{n}\sum_{l = 1}^q \epsilon_{k,i,l}^{(1)}\epsilon_{k,j,l}^{(2)} \right \} = 
\sqrt{r_{i,i}^{(1)}r_{j,j}^{(2)}} \; \rho_{S_{1,2},i,j} \; \text{tr}(\Rho_{T_{1,2}})/q = \sqrt{r_{i,i}^{(1)}r_{j,j}^{(2)}} \; \varrho_{i,j}^{(1,2)}.
\end{eqnarray*}
Therefore we can estimate $\varrho_{i,j}^{(1,2)}$ by 
\begin{eqnarray} \label{eq:est_rhoS12ij}
\hat{\varrho}_{i,j}^{(1,2)} &=& \hat{\Cov} (\epsilon_{\cdot,i,\cdot}^{(1)}, \epsilon_{\cdot,j,\cdot}^{(2)})/\sqrt{\hat{r}_{i,i}^{(t)}\hat{r}_{j,j}^{(t)}},
\quad \textrm{ and } \quad
\hat{\Cov} (\epsilon_{\cdot,i,\cdot}^{(1)}, \epsilon_{\cdot,j,\cdot}^{(2)}) = \frac{1}{nq} \sum_{k = 1}^{n}\sum_{l = 1}^q \heps_{k,i,l}^{(1)}\heps_{k,j,l}^{(2)}.
\end{eqnarray}
Correspondingly, when $\S_{T_t}$, $\S_{T_{1,2}}$ and thus $\Rho_{T_{1,2}}$ are known, we can estimate $\Theta_{i,j}$ by  
\begin{eqnarray} \label{eq:est_Thetaij}
\hat{\Theta}_{i,j} = \hat{\theta}_{i,j}^{(1)} + \hat{\theta}_{i,j}^{(2)} - \frac{2}{nq} \left\{ \hat{\varrho}_{i,i}^{(1,2)}\hat{\varrho}_{j,j}^{(1,2)}+\hat{\varrho}_{i,j}^{(1,2)}\hat{\varrho}_{j,i}^{(1,2)} \right\} \frac{q\|\Rho_{T_{1,2}}\|_F^2 }{ \text{tr}(\Rho_{T_{1,2}})^2}.
\end{eqnarray}
We show in Section \ref{sec:theory:results} that $\hat{\Theta}_{i,j}$ in \eqref{eq:est_Thetaij} provides an accurate estimation of $\Theta_{i,j}$, with an error rate of order $o_p(1/ \log p)$, when $\S_{T_t}$ and $\S_{T_{1,2}}$ are known.

When $\S_{T_t}$ and $\S_{T_{1,2}}$ are unknown, we first estimate $\Rho_{T_{1,2}}$ by
\begin{equation}  \label{eq:est_tildeT}
\hat{\Rho}_{T_{1,2}}^{(d)} = \frac{1}{np} \sum_{k = 1}^n\sum_{i = 1}^p \left[ \left\{ \Y^{(1,d)}_{k,i,\cdot} - \frac{1}{np}\sum_{k = 1}^n\sum_{i = 1}^p \Y^{(1,d)}_{k,i, \cdot} \right\}^\T \left\{ \Y^{(2,d)}_{k,i, \cdot} - \frac{1}{np}\sum_{k = 1}^n \sum_{i = 1}^p \Y^{(2,d)}_{k,i, \cdot} \right\} \right],
\end{equation}
where $\Y^{(t,d)}_{k,i,\cdot}$ is the $i$th row of $\Y^{(t,d)}_{k} = \X^{(t)}_{k}{\hat\S}_{T_t}^{-1/2}$, and $\hat\S_{T_t}$ is an estimator of $\S_{T_t}$. We then plug \eqref{eq:est_tildeT} into \eqref{eq:est_Thetaij}. Again we show in Section \ref{sec:theory:results} that this estimator also provides an accurate estimation of $\Theta_{i,j}$, with an error rate of order $o_p(1/ \log p)$, when $\S_{T_t}$ and $\S_{T_{1,2}}$ are unknown. 

We make a few remarks about our proposed variance correction. First, a crucial component of our method is to pool data information of the $p$-dimensional spatial and $q$-dimensional temporal measurements of $n$ subjects in our estimations. The data pooling is possible due to the facts that $\EE \{ \Y^{(1)} \cdot (\Y^{(2)})^{\T} \} = \text{tr}(\Rho_{T_{1,2}}) \cdot \S_{S_{1,2}}$ and $\EE \{ (\Y^{(1)})^{\T} \cdot (\Y^{(2)}) \} = \text{tr}(\S_{S_{1,2}}) \cdot \Rho_{T_{1,2}}$. Consequently, we can pool the columns of $\Y^{(t)}$ to estimate $\S_{S_{1,2}}$, and the rows of $\Y^{(t)}$ to estimate $\Rho_{T_{1,2}}$, up to a constant. More specifically, when $\S_{T_t}$ and $\S_{T_{1,2 }}$ are known, we  pool $2nq$ samples to estimate the within-sample variance as in \eqref{theij}, and the between-sample spatial dependency as in \eqref{eq:est_rhoS12ij} and \eqref{eq:est_Thetaij}. When $\S_{T_t}$ and $\S_{T_{1,2 }}$ are unknown, we also pool $2np$ samples to obtain the estimates $\hat\S_{T_t}^{-1/2}, t=1,2$, and estimate the temporal dependency between the before-stimulus scan and the after-stimulus scan as in \eqref{eq:est_tildeT}. Such data pooling is the main difference between our method and a naive solution, which estimates the dependency between the paired samples by the usual sample covariance, namely, estimating $\Cov \left\{ \epsilon_{k,i,l_1}^{(1)},\epsilon_{k,j,l_2}^{(2)} \right\}$ by $n^{-1}\sum_{k=1}^n\hat\epsilon_{k,i,l_1}^{(1)}\hat\epsilon_{k,j,l_2}^{(2)}$, for each $1\leq i<j\leq p, 1\leq l_1,l_2\leq q$. Note that, the latter approach only uses $n$ observations to estimate the dependence structure without any data pooling, and as a result, it can not guarantee the estimation error rate required to ensure the performance of the test. 

Second, we note that the spatial and temporal covariances $\S_{S_{1,2}}$ and $\Rho_{T_{1,2}}$ are only identifiable up to a constant. However, this does not affect our test statistic, nor our variance estimation. This is because, when replacing $(\S_{S_{1,2}}, \Rho_{T_{1,2}})$ with $(c\S_{S_{1,2}}, \Rho_{T_{1,2}}/c)$, where $c$ is any positive factor, the terms $\varrho_{i,j}^{(1,2)}$ and $\|\Rho_{T_{1,2}}\|_F^2 / \text{tr}(\Rho_{T_{1,2}})^2$ remain the same, in which the factor $c$ is cancelled.

Third, \cite{chen2018testing} developed a variance correction method for matrix-valued data, but for a single group of samples. In contrast, we perform variance correction for two stages of samples from the same population. We first separate the spatial and temporal structures,  so that the resulting test statistics do not require variance correction within each sample. Our variance correction differs from that of \cite{chen2018testing} considerably. On the other hand, if the temporal covariance between two stages has some particular structure, e.g., if it is sparse, then the method of \cite{chen2018testing} may be applied to our procedure, by thresholding $\hat\Rho_{T_{1,2}}^{(d)}$ in \eqref{eq:est_tildeT} accordingly. In this paper, however, we do not impose any structural condition on the temporal dependence, and thus we use the general sample covariance estimator in \eqref{eq:est_tildeT} instead.

\subsection{Multiple testing}
\label{sec:methods:multi}

We next develop a multiple testing procedure for $H_{0,i,j}: \rho_{S_{1},i,j}= \rho_{S_{2},i,j}, 1 \leq i < j \leq p$, so to identify spatial locations with their conditional dependence changed before and after the stimulus. With a total of $p(p-1)/2$ simultaneous tests, the key is to control false discovery. Let $h$ be the rejection threshold value such that $H_{0, i, j}$ is rejected if $|W_{i,j}| \geq h$, and $\mathcal{H}_0 := \{(i, j): \rho_{S_1, i,j} = \rho_{S_2, i, j}, 1 \leq i < j \leq p\}$ be the set of true nulls. Then the false discovery proportion (FDP) and the false discovery rate (FDR) are computed as
\begin{equation*}
\text{FDP}(h)=\frac{\sum_{(i,j)\in \mathcal{H}_0}I(|W_{i,j}|\geq h)}{\sum_{1\leq i<j\leq p}I(|W_{i,j}|\geq h)\vee 1}, \quad  \text{FDR}(h)=\EE\{\text{FDP}(h)\}.
\end{equation*}

Our multiple testing procedure is based on the test statistic $W_{i,j}$ derived in Section \ref{sec:method:test_stat}, with the corrected variance estimates $\hat\Theta_{i,j}$ derived in Section \ref{sec:methods:var_correct}. The rest of the procedure is similar to that of the two-sample independent test of \citet{xia2018two}. We thus only outline the main steps here. First, we compute the paired-test statistics $W_{i,j}$ in \eqref{eq:test_W} for all $1 \leq i < j \leq p$. Next we estimate the false discovery proportion by 
\begin{eqnarray*} \label{eq:multiple_fdr}
\hat{FDP}(h) = \frac{2\{1 - \Phi(h)\}(p^2 - p)/2}{\sum_{1 \leq i < j \leq p}I(|W_{i, j}| \geq h) \lor 1},
\end{eqnarray*}
where $\Phi(\cdot)$ is the standard normal cumulative distribution function. Here we conservatively estimate $|\mathcal{H}_0|$ by $(p^2-p)/2$, as it is at maximum $(p^2-p)/2$ and is close to $(p^2-p)/2$ when $\R_{S_1}-\R_{S_2}$ is sparse. Next, we compute the rejection threshold value $\hat{h}_\alpha$ under a given significance level $\alpha$ as 
\begin{eqnarray} \label{eq:sig_alpha}
\hat{h}_{\alpha} = \inf \left \{0 \leq h \leq 2 (\log p)^{1/2}: \hat{FDP}(h) \leq \alpha \right \}.
\end{eqnarray}
If $\hat{h}_{\alpha}$ does not exist, we set $\hat{h}_{\alpha} = 2 (\log p)^{1/2}$. Finally, we reject $H_{0,i,j}$ if and only if $|W_{i,j}| \geq \hat{h}_{\alpha}$ for each $1 \leq i < j \leq p$. In Section \ref{sec:theory:results} we show that the above multiple testing procedure can control FDR at the pre-specified level asymptotically.

\section{Theory}
\label{sec:theory:results}

We study in this section the asymptotic properties of the proposed testing procedure. In the interest of space, we present all the regularity conditions (A1)-(A7) in the appendix. We first show that the corrected variance estimator of $\Theta_{i,j}$ we develop in Section \ref{sec:methods:var_correct} achieves the estimation error rate of $o_p(1/ \log p)$. We then show that, based on such an error rate, the subsequent multiple testing procedure can control the false discovery asymptotically. 

When $\S_{T_t}$ and $\S_{T_{1,2}}$ are known, our variance estimator is $\hat\Theta_{i,j}$ as given in \eqref{eq:est_Thetaij}. The next proposition establishes its error rate. 

\begin{proposition}\label{lemma:corr_est}
Suppose (A1), (A3) and (A5) hold. Then we have
\begin{eqnarray*}
\underset{i,j}{\max}|nq(\hat{\Theta}_{i,j} - \Theta_{i,j})| = o_p(1/\log p).
\end{eqnarray*}
\end{proposition}

When $\S_{T_t}$ and $\S_{T_{1,2}}$ are unknown, we denote our variance estimator as $\hat{\Theta}_{i,j}^{(d)}$, which is obtained by plugging the estimator $\hat{\Rho}_{T_{1,2}}^{(d)}$ in \eqref{eq:est_tildeT} into \eqref{eq:est_Thetaij}. The next proposition establishes its error rate. 

\begin{proposition}\label{lemma:data}
Suppose (A1), (A3) and (A6)-(A7) hold. Then we have 
\begin{equation*}
\underset{i,j}{\max}|nq(\hat{\Theta}_{i,j}^{(d)} - \Theta_{i,j})| = o_p(1/\log p).
\end{equation*}
\end{proposition}

\noindent
The above two propositions show that, the variance estimation error is bounded by the same error rate asymptotically, when $\S_{T_t}$ and $\S_{T_{1,2}}$ are unknown and when they are known.

The next theorem shows that, for the dependent samples, as long as the majority of the regression residuals are not highly correlated with each other under the null hypothesis, then the FDR can be controlled asymptotically at the pre-specified level $\alpha$ following the multiple testing procedure outlined in Section \ref{sec:methods:multi}. 

\begin{theorem}\label{thm:fdr_control}
Let $\ell_0 = |\mathcal{H}_0|$ and $\ell = (p^2 - p)/2$. Suppose $\ell_0 \geq \tilde{c}_0 p^2$ for some constant $\tilde{c}_0 > 0$, and $p \leq \tilde{c}_1(nq)^{\tilde{c}_2}$ for some $\tilde{c}_1, \tilde{c}_2 > 0$. Let $\hat{h}_{\alpha}$ denote the threshold value in \eqref{eq:sig_alpha}. Then, when (A1)-(A5) hold and $\S_{T_t}$ and $\S_{T_{1,2}}$ are  known, or when (A1)-(A4), (A6)-(A7) hold and $\S_{T_t}$ and $\S_{T_{1,2}}$ are unknown, we have  
\begin{eqnarray*}
\frac{FDR(\hat h_{\alpha})}{\alpha \ell_0/\ell} {\rightarrow} 1,~~~ \frac{FDP(\hat h_{\alpha})}{\alpha \ell_0/\ell} \overset{p}{\rightarrow} 1,~~~ \text{as } (nq, p) \rightarrow \infty.
\end{eqnarray*}
\end{theorem}

In addition to false discovery control, the asymptotic power analysis is another interesting problem. It relies on the specific structure of the connectivity network. In Section \ref{sec:simulation}, we conduct extensive simulations to study the power of our test under numerous network structures, and we leave the theoretical power analysis as future research.

\section{Simulations}
\label{sec:simulation}

\subsection{Empirical FDR and power with and without variance correction}
\label{sec:normal}

We conduct numerous simulations to study the finite sample performance of our proposed variance-corrected testing procedure. We also compare with the two-sample test of \citet{xia2018two}, which ignores the correlation before and after the stimulus and does not correct the variance accordingly. In all the simulations, we use Lasso to estimate the regression coefficient $\beta_{i}^{(t)}$, and use the banded covariance approach to estimate $\S_{T_t}$. We set the FDR level at $\alpha = 1\%$. 

We examine a set of spatial and temporal dimensions, $(p, q) \in \{(200, 50), (200, 200), (800, 200)\}$, while we fix the sample size at $n=15$. We consider two temporal covariance structures: an autoregressive structure, where $\S_{T_t} = (\sigma_{T_t, i, j})$, $\sigma_{T_t, i, j} = 0.4^{|i -j|}$ if $t=1$, and $\sigma_{T_t, i, j} = 0.5^{|i - j|}$ if $t=2$, $1 \leq i, j \leq p$, and a moving average structure, where $\S_{T_t} = (\sigma_{T_t, i, j})$, $\sigma_{T_t, i,j} = 1/(|i - j| + 1)$ for $|i-j| < 3$ if $t = 1$, and $\sigma_{T_t, i, j} = 1/(|i - j| + 1)$ for $|i - j| \leq 4$ if $t=2$. We also consider three spatial covariance structures: a banded graph, with bandwidth equal to $3$ (\citealp{zhao2012huge}), a hub graph, with rows and columns evenly partitioned into 20 disjoint groups, and a small-world graph, with $5$ starting neighbors and $5\%$ probability of rewiring (\citealp{van2016ridge}). We first generate $\O_{S_1}$ according to one of the above spatial structures, then construct $\O_{S_2}$ by randomly eliminating $50 \%$ percent of the edges of $\O_{S_1}$. 

Moreover, we consider two settings of correlation patterns before and after the stimulus. In Setting I, we set $\S_{S_{1,2}} = \gamma \S_{S_1}$, where $\gamma$ is the overall correlation level and $|\gamma| \leq 1$. Since $\gamma$ plays its role through $\gamma^2$, its sign does not matter, and we choose $\gamma \in \{0, 0.2, 0.4, 0.6\}$. When $\gamma = 0$, it reduces to the two-sample independent case, whereas a larger value of $\gamma$ implies a stronger before-and-after stimulus correlation. We next set $\Rho_{T_{1,2}}$ as a diagonal matrix with $\Rho_{T_{1,2}, i,i} = -1$ if $i \equiv k \pmod{15}, k \in \{1, 3, 5\}$, and $1$ otherwise. Here for three positive integers $a$, $b$ and $c$, $a \equiv b \; (\text{mod}~c)$ means that, when divided by $c$, $a$ and $b$ have the same remainder that is non-negative and smaller than $c$. In this setting, it follows that
\begin{equation*}
  (\rho_{S_{1,2},i,i}^{(1,2)}\rho_{S_{1,2},j,j}^{(1,2)}+\rho_{S_{1,2},i,j}^{(1,2)}\rho_{S_{1,2},j,i}^{(1,2)}) = \gamma^2 \sqrt{r_{i,i}^{(1)}r_{i,i}^{(2)}r_{j,j}^{(1)}r_{j,j}^{(2)}} (\omega_{S_2, i, i}\omega_{S_2, j, j} + \omega_{S_2, i, j}\omega_{S_2, j, i}) > 0,
\end{equation*}
as long as $\gamma > 0$, where we utilize the facts that $\Sigma_{S_{1,2}} = \gamma \Sigma_{S_1} = \gamma \Omega_{S_1}^{-1}$, and $\Omega_{S_2}$ is a positive definitive matrix. Correspondingly, $\Theta_{i,j}$ is smaller than that of the independent case, and the test statistic $W_{i,j}$ would be larger than that without variance correction in its absolute value. For this setting, the two-sample test without variance correction is to yield a smaller power, as it is more conservative in rejecting the null hypothesis in this setting. 

In Setting II, we set $\Rho_{T_{1,2}}$ in the same way, but set $\S_{S_{1,2}, i, j} = \gamma \cdot \S_{S_1,i,j}  (-1)^{i + j}$, if $i \neq j$, and $\gamma \cdot \S_{S_1, i,i} (1 - 2 \cdot \mathbf{1}[ i \equiv k \pmod{7}, k \in \{1, 3, 5\}] )$ if $i = j$, where $\mathbf{1}(\cdot)$ is the indicator function. In this setting, we no longer have a simplified expression for $\rho_{S_{1,2},i,i}^{(1,2)}\rho_{S_{1,2},j,j}^{(1,2)}+\rho_{S_{1,2},i,j}^{(1,2)}\rho_{S_{1,2},j,i}^{(1,2)}$, but empirically, we have observed that this term is negative for about half of $(i,j)$ pairs regardless of the choice of the spatial structure and the dimension $p$. For those pairs, $\Theta_{i,j}$ is larger than that of the independent case, and the test statistic $W_{i,j}$ would be smaller than that without variance correction in its absolute value. For this setting, the two-sample test without variance correction is to yield an overestimated FDR in this setting. 

Tables \ref{tbl:general_up} and \ref{tbl:general_down} report the empirical FDR and the empirical power, both in percentage, out of $100$ data replications for the two settings, respectively. We make the following observations.

For Setting I, when $(p, q) = (200, 50)$, the test with variance correction controls the FDR around the anticipated level of $\alpha = 1\%$, whereas the test without variance correction yields a much lower FDR than the significance level. Moreover, as the correlation strength $\gamma$ increases, the power of the test with variance correction improves considerably compared to the test without correction. Similar qualitative patterns are observed for $(p, q) = (200, 200)$ and $(p, q) = (800, 200)$.

For Setting II, for different combinations of $(p, q)$ and spatial structures, the test with variance correction again controls the FDR close to the significance level, while the test without correction fails to control FDR as $\gamma$ increases. When $(p, q) = (200, 50)$, the test with correction is slightly inferior to that without correction for the banded graph in terms of power. This is not surprising though, as it is attributed to the inflated FDR. For other spatial structures, the test with correction clearly outperforms the one without correction. When $(p, q) = (200, 200)$, we observe that the power of both tests increases to $100\%$ or close. For FDR, the inflation issue still remains for the test without correction. When $(p, q) = (800, 200)$, we observe a similar qualitative pattern.

In summary, our proposed test with variance correction can control the false discovery and attain a good power for a range of strength of correlation before and after the stimulus. By contrast, the test without correction has inferior power performance for Setting I, and fails to control the FDR and yields an inflated power for Setting II as this correlation increases. We also report the mean squared error of $\hat{\Theta}^{(d)}$ in Section \ref{appendix:theta_mse} of the appendix.

\subsection{Sensitivity analysis}
\label{sec:sensitivity}

We next carry out sensitivity analysis to evaluate the performance of our test when the data deviates from the matrix normal distribution. We first replace the normal distribution by a $t$ distribution. Specifically, we follow the data generation mechanism as before, while we set $\S_{S_{1,2}}$ as a diagonal matrix with $\S_{S_{1,2},i,i} = \gamma \cdot \S_{S_1, i,i} (1 - 2 \cdot \mathbf{1}[ i \equiv k \pmod{7}, k \in \{1, 3, 5\}])$ and $\gamma = 0.6$. Since a normal random vector $\X \sim N(\bold 0, \S)$ can be represented as $\X = \S^{1/2} \Z$, where $\Z \sim N(\bold 0, I_p)$, we replace the Gaussian entries in $\Z$ with $t$-distributed random variables with degree of freedom $df \in \{4, 6, 8\}$. We report the empirical FDR and power out of 100 data replications in Table \ref{tbl:sensitivity}, part I. It is seen that, our test manages to control the FDR reasonably well, and attains a good power under different dependence structures.
    
We then examine the performance of our method with regard to the off-diagonal Kronecker product structure, i.e,  $\Cov (\text{vec}\{\bold X^{(1)}\}, \text{vec}\{\bold X^{(2)}\}) = \S_{S_{1,2}} \otimes \S_{T_{1,2}}$. We again follow the data generation mechanism as before, but set $\S_{S_{1,2}}$ as a diagonal matrix with $\S_{S_{1,2},i,i} = \gamma \cdot \S_{S_1, i,i} (1 - 2 \cdot \mathbf{1}[ i \equiv k \pmod{7}, k \in \{1, 3, 5\}])$, $\gamma = 0.6$, and $(p, q) = (200, 50)$. We further perturb $\S_{1, 2} = \S_{S_{1,2}} \otimes \S_{T_{1,2}}$ in two steps: we randomly sample $p^\star \%$ entries of $\S_{1, 2}$, where $p^\star \in \{0, 1, 5, 10\}$, then replace those entries with i.i.d.\ Gaussian random variables of mean zero and standard deviation $\nu$, where $\nu = l^\star \times \text{the magnitude for the entries of } \S_{1,2}$, and $l^\star \in \{0.1, 1\}$. We report the empirical FDR and power out of 100 data replications in Table \ref{tbl:sensitivity}, part II. It is seen that, our test maintains a reasonably good performance in this setup too.

These results show that our method is relatively robust with regard to the joint matrix normal assumption \eqref{eq:joint_dist_X}. We also comment that, it is possible to extend our test to semiparametric normal copula setting. \cite{liu2012high} and \cite{xue2012regularized} studied the vector-valued case. Following a similar idea of marginal monotonic transformation, it is possible to develop a paired test in the matrix-valued setting. We leave the full investigation as future research.

\section{Alzheimer's disease data analysis}
\label{sec:realdata}
Alzheimer's disease (AD) is an irreversible neurodegenerative disorder, and is characterized by progressive impairment of cognitive and memory functions. It is the leading form of dementia in the elderly subjects. With the aging of the worldwide population, the number of affected people is rapidly increasing and is projected to be $13.8$ million in the United States, and 1 in 85 worldwide by year 2050 (\citealp{Brookmeyer2007, brookmeyer2011national}). It thus has become an international imperative to understand, diagnose, and treat this disorder. Accumulated evidences have suggested that alterations in brain connectivity networks are predictive of cognitive function and decline, and hold crucial insights about the disease pathology of AD (\citealp{fox2010clinical}). 

We analyzed a dataset from the Alzheimer's Disease Neuroimaing Initiative (ADNI). ADNI is an ongoing, longitudinal, multi-center study designed to develop clinical, imaging, genetic, and biochemical biomarkers for the early detection and tracking of AD. We focused on 23 subjects from ADNI who experienced conversion from mild cognitive impairment (MCI), a prodromal stage of AD, to AD during the 24-month follow-up. The primary scientific question of interest is to investigate the change of brain connectivity patterns before and after the conversion. All fMRI scans were  resting-state and preprocessed, including slice timing correction, motion correction, spatial smoothing, denoising by regressing out motion parameters, white matter and cerebrospinal fluid time courses, and band-pass filtering. The data were then aligned and parcellated using the Anatomical Automatic Labeling atlas (\citealp{tzourio2002automated}). The resulting data is a region by time matrix for each subject, with the spatial dimension $p = 116$ and the temporal dimension $q = 130$.

\begin{figure}[t!]
\centering
\includegraphics[width=\textwidth]{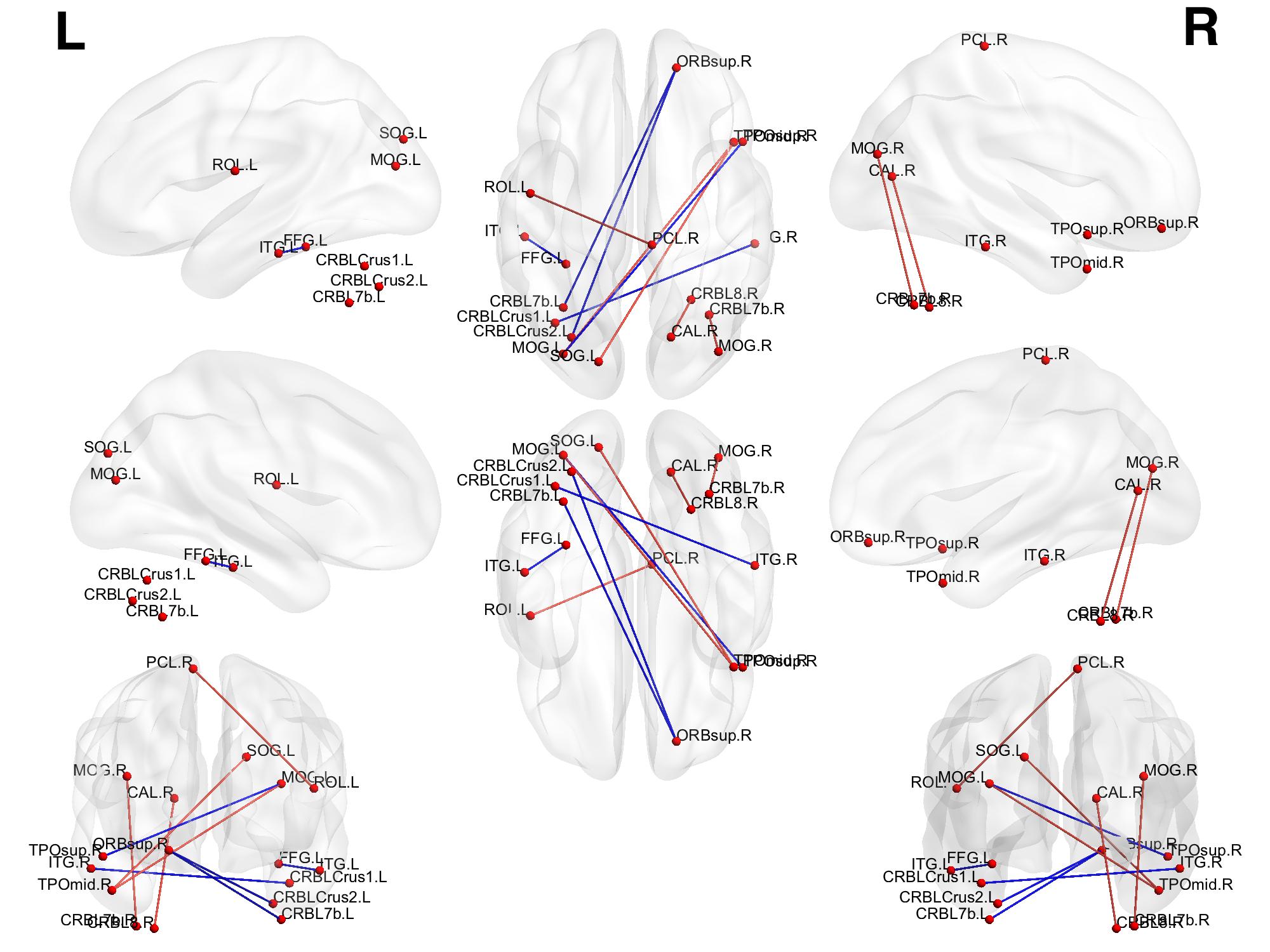}
\caption{Top 10 differentiating links of the brain connectivity networks of the 23 subjects of the ADNI database before and after the conversion from MCI to AD. All the associated $p$-values are smaller than $1e-13$. Red links are those enhanced after the conversion, and blue links are those weakened after the conversion.}
\label{fig:brainlinks}
\end{figure}

We first examined the quantile-quantile plot, which shows no clear deviation from the normal distribution. We next applied the testing procedure of \cite{aston2017tests} to test if the data conforms with the Kronecker product structure. The p-values of the test before and after the conversion are $0.17$ and $0.11$, respectively, which suggests that the product structure seems to reasonably hold for this dataset. We then applied our proposed variance-corrected testing procedure to this data. In our analysis, we did not correct for potential confounder effects, but our test can be equally applied to the corrected data.  Figure \ref{fig:brainlinks} plots those top differentiating links whose corresponding $p$-values are smaller than $1e-13$, and the associated brain regions visualized with the BrainNet Viewer (\citealp{Xia2013}). Table \ref{tbl:top30_real} further reports the top $30$ links that were found different before and after the conversion, with their associated $p$-values and the directions of the change. It is seen that the differentiating links concentrate on the cerebellum. The cerebellum is critical in the distributed neural circuits subserving not only motor function but also autonomic, limbic and cognitive behaviors. There is recently increased interest in exploring the role of the cerebellum in neurodegenerative disorders, in particular Alzheimer's disease (\citealp{Jacobs2018}). Our findings provide a useful support to the existing literature.

\section{Supplementary material}\label{sec:supp_material}
The computer code in \textit{R} for the simulation and data analysis can be found at \url{https://github.com/Elric2718/PairedTestPrecisionMatrix} with the commit number 631c4e4. The Alzheimer's disease dataset can be found at \url{https://doi.org/10.6084/m9.figshare.9643010.v3}.

\section*{Acknowledgments}
The authors thank the Associate Editor and two anonymous reviewers of the \textit{Biostatistics} journal for constructive suggestions and questions that helped improve the manuscript. Conflict of Interest: None declared.

\section*{Funding}
Xia's research was partially supported by NSFC grants 11771094, 11690013 and The Recruitment Program of Global Experts Youth Project. Li's research was partially supported by NSF grant DMS-1613137 and NIH grants R01AG034570 and R01AG061303.


\newsavebox{\tablebox}
\begin{sidewaystable}[hptb]\small\addtolength{\tabcolsep}{-4pt}
\begin{center}
\caption{The empirical FDR (with standard error in the parenthesis) and the empirical power, both in percentage, for Setting I.}        
\begin{lrbox}{\tablebox}
\begin{tabular}{c@{\hspace{1em}} |c@{\hspace{1em}}  |r@{\hspace{1em}} r@{\hspace{1em}}|r@{\hspace{1em}} r @{\hspace{1em}}|r@{\hspace{1em}}r@{\hspace{1em}} |r@{\hspace{1em}} r@{\hspace{1em}}|r@{\hspace{1em}} r @{\hspace{1em}}|r@{\hspace{1em}}r@{\hspace{1em}} }
\hline
\multicolumn{2}{c|}{temporal structure}&\multicolumn{6}{c}{moving average}&\multicolumn{6}{|c}{autoregressive}\\[0pt]
\hline
\multicolumn{2}{c|}{spatial structure} &\multicolumn{2}{c}{banded}&\multicolumn{2}{|c}{hub}&\multicolumn{2}{|c}{small}&\multicolumn{2}{|c}{banded}&\multicolumn{2}{|c}{hub}&\multicolumn{2}{|c}{small}\\  [0pt]
\hline
\multicolumn{2}{c|}{variance correction}&\multicolumn{1}{c|}{\cmark}&\multicolumn{1}{c|}{\xmark}&\multicolumn{1}{c|}{\cmark}&\multicolumn{1}{c|}{\xmark}&\multicolumn{1}{c|}{\cmark}&\multicolumn{1}{c|}{\xmark}&\multicolumn{1}{c|}{\cmark}&\multicolumn{1}{c|}{\xmark}&\multicolumn{1}{c|}{\cmark}&\multicolumn{1}{c|}{\xmark}&\multicolumn{1}{c|}{\cmark}&\multicolumn{1}{c|}{\xmark}\\[0pt]
\hline\hline\hline
$(p,q)$  & $\gamma$ &\multicolumn{12}{c}{Empirical FDR (SE)  (in \%)}\\[0pt]
\hline
\multirow{5}{*}{$(200,50)$}  
&$0.0$ &0.7(0.5)& 0.7(0.5)  & 0.8(1.2)  & 0.7(1.2) & 1.1(1.3) & 1.1(1.2) & 0.8(0.5) & 0.8(0.5) & 0.8(1.1) & 0.8(1.1)  & 1.0(1.2)  & 0.9(1.2) \\[0pt]
&$0.2$ &0.8(0.6)& 0.5(0.5)  & 1.1(1.3)  & 0.7(1.1) & 1.1(1.1) & 0.8(0.9) & 0.7(0.4) & 0.5(0.4) & 1.0(1.5) & 0.7(1.4)  & 1.0(1.0)  & 0.7(0.8) \\[0pt]
&$0.4$ &0.9(0.6)& 0.1(0.2)  & 1.0(1.2)  & 0.1(0.5) & 1.2(1.1) & 0.3(0.6) & 0.7(0.6) & 0.1(0.2) & 1.0(1.1) & 0.2(0.6)  & 1.1(1.1)  & 0.3(0.7) \\[0pt]
&$0.6$ &0.9(0.6)& 0.0(0.1)  & 1.1(1.2)  & 0.1(0.5) & 1.3(0.8) & 0.1(0.3) & 0.9(0.6) & 0.0(0.1) & 1.1(1.3) & 0.0(0.3)  & 1.3(1.0)  & 0.1(0.3) \\[0pt]
\hline
\multirow{5}{*}{$(200,200)$} 
&$0.0$ &0.8(0.5)& 0.8(0.5)  & 1.1(1.2)  & 1.1(1.2) & 0.7(0.4) & 0.7(0.4) & 0.9(0.5) & 0.9(0.5) & 0.9(0.9) & 0.8(0.9)  & 0.9(0.5)  & 0.9(0.4) \\[0pt]
&$0.2$ &0.9(0.6)& 0.5(0.5)  & 1.1(1.3)  & 0.8(1.0) & 0.9(0.4) & 0.6(0.4) & 0.8(0.5) & 0.5(0.4) & 1.2(1.2) & 0.7(0.9)  & 0.9(0.4)  & 0.5(0.3) \\[0pt]
&$0.4$ &1.1(0.7)& 0.2(0.2)  & 1.0(1.0)  & 0.2(0.4) & 0.8(0.4) & 0.2(0.2) & 0.9(0.5) & 0.2(0.2) & 1.1(1.0) & 0.3(0.5)  & 0.8(0.5)  & 0.2(0.2) \\[0pt]
&$0.6$ &1.1(0.6)& 0.0(0.1)  & 1.1(1.2)  & 0.0(0.2) & 1.0(0.4) & 0.0(0.1) & 1.0(0.6) & 0.0(0.1) & 1.5(1.2) & 0.0(0.2)  & 0.9(0.5)  & 0.0(0.1) \\[0pt]
\hline
\multirow{5}{*}{$(800,200)$} 
&$0.0$ &0.7(0.2)& 0.7(0.2)  & 0.9(0.7)  & 0.9(0.6) & 0.7(0.2) & 0.7(0.2) & 0.7(0.2) & 0.7(0.2) & 0.9(0.6) & 0.9(0.6)  & 0.8(0.2)  & 0.8(0.2) \\[0pt]
&$0.2$ &0.8(0.3)& 0.5(0.2)  & 0.9(0.6)  & 0.5(0.5) & 0.8(0.2) & 0.5(0.2) & 0.7(0.3) & 0.4(0.2) & 0.9(0.7) & 0.5(0.5)  & 0.8(0.2)  & 0.5(0.1) \\[0pt]
&$0.4$ &0.9(0.3)& 0.1(0.1)  & 1.1(0.7)  & 0.3(0.4) & 0.8(0.2) & 0.1(0.1) & 0.8(0.3) & 0.1(0.1) & 0.8(0.6) & 0.3(0.4)  & 0.8(0.2)  & 0.1(0.1) \\[0pt]
&$0.6$ &0.9(0.3)& 0.0(0.1)  & 1.1(0.8)  & 0.2(0.3) & 0.9(0.2) & 0.0(0.0) & 0.9(0.3) & 0.0(0.0) & 1.1(0.6) & 0.3(0.4)  & 0.9(0.2)  & 0.0(0.0) \\[0pt]
\hline
& &\multicolumn{12}{c}{Empirical power (in \%)}\\[0pt]
\hline
\multirow{5}{*}{$(200,50)$}  
&$0.0$ &90.7& 90.8  & 54.8  & 54.9 & 19.4 & 19.4 & 90.2 & 90.2 & 54.9 & 54.9  & 19.3  & 19.4 \\[0pt]
&$0.2$ &93.0& 91.2  & 58.6  & 55.4 & 22.8 & 18.9 & 92.8 & 91.0 & 55.2 & 51.1  & 23.8  & 19.7 \\[0pt]
&$0.4$ &96.8& 91.2  & 69.8  & 53.8 & 33.2 & 17.0 & 97.2 & 91.9 & 68.7 & 53.1  & 31.6  & 16.0 \\[0pt]
&$0.6$ &99.1& 92.4  & 85.3  & 43.0 & 45.6 & 14.8 & 99.3 & 93.4 & 86.4 & 42.4  & 46.5  & 15.5 \\[0pt]     
\hline
\multirow{5}{*}{$(200,200)$}
&$0.0$ &100.0& 100.0  & 100.0  & 100.0 & 99.9 & 99.9 & 100.0 & 100.0 & 100.0 & 100.0  & 99.9  & 99.9 \\[0pt]
&$0.2$ &100.0& 100.0  & 100.0  & 100.0 & 99.9 & 99.9 & 100.0 & 100.0 & 100.0 & 100.0  & 100.0  & 99.9 \\[0pt]
&$0.4$ &100.0& 100.0  & 100.0  & 100.0 & 100.0 & 100.0 & 100.0 & 100.0 & 100.0 & 100.0  & 100.0  & 100.0 \\[0pt]
&$0.6$ &100.0& 100.0  & 100.0  & 100.0 & 100.0 & 100.0 & 100.0 & 100.0 & 100.0 & 100.0  & 100.0  & 100.0 \\[0pt]     
\hline
\multirow{5}{*}{$(800,200)$}
&$0.0$ &100.0& 100.0  & 62.0  & 62.2 & 99.6 & 99.6 & 100.0 & 100.0 & 61.1 & 61.4  & 99.5  & 99.5 \\[0pt]
&$0.2$ &100.0& 100.0  & 60.3  & 60.6 & 99.8 & 99.7 & 100.0 & 100.0 & 60.1 & 60.5  & 99.8  & 99.7 \\[0pt]
&$0.4$ &100.0& 100.0  & 57.7  & 50.1 & 100.0 & 99.8 & 100.0 & 100.0 & 57.7 & 50.2  & 99.9  & 99.8 \\[0pt]
&$0.6$ &100.0& 100.0  & 52.2  & 47.1 & 100.0 & 99.8 & 100.0 & 100.0 & 54.7 & 47.2  & 100.0  & 99.8 \\[0pt]     
\hline
\end{tabular}
\end{lrbox}
\scalebox{1.0}{\usebox{\tablebox}}
\label{tbl:general_up}
\end{center}
\end{sidewaystable}

\begin{sidewaystable}[hptb]\small\addtolength{\tabcolsep}{-4pt}
\begin{center}
\caption{The empirical FDR (with standard error in the parenthesis) and the empirical power, both in percentage, for Setting II.}    
\begin{lrbox}{\tablebox}
\begin{tabular}{c@{\hspace{1em}} |c@{\hspace{1em}}  |r@{\hspace{1em}} r@{\hspace{1em}}|r@{\hspace{1em}} r @{\hspace{1em}}|r@{\hspace{1em}}r@{\hspace{1em}} |r@{\hspace{1em}} r@{\hspace{1em}}|r@{\hspace{1em}} r @{\hspace{1em}}|r@{\hspace{1em}}r@{\hspace{1em}} }
\hline
\multicolumn{2}{c|}{temporal structure}&\multicolumn{6}{c}{moving average}&\multicolumn{6}{|c}{autoregressive}\\[0pt]
\hline
\multicolumn{2}{c|}{spatial structure} &\multicolumn{2}{c}{banded}&\multicolumn{2}{|c}{hub}&\multicolumn{2}{|c}{small}&\multicolumn{2}{|c}{banded}&\multicolumn{2}{|c}{hub}&\multicolumn{2}{|c}{small}\\  [0pt]
\hline
\multicolumn{2}{c|}{variance correction}&\multicolumn{1}{c|}{\cmark}&\multicolumn{1}{c|}{\xmark}&\multicolumn{1}{c|}{\cmark}&\multicolumn{1}{c|}{\xmark}&\multicolumn{1}{c|}{\cmark}&\multicolumn{1}{c|}{\xmark}&\multicolumn{1}{c|}{\cmark}&\multicolumn{1}{c|}{\xmark}&\multicolumn{1}{c|}{\cmark}&\multicolumn{1}{c|}{\xmark}&\multicolumn{1}{c|}{\cmark}&\multicolumn{1}{c|}{\xmark}\\[0pt]
\hline\hline\hline
$(p,q)$  & $\gamma$ &\multicolumn{12}{c}{Empirical FDR (SE)  (in \%)}\\[0pt]
\hline
\multirow{5}{*}{$(200,50)$}  
&$0.0$ &0.7(0.6)& 0.7(0.6)  & 0.5(1.2)  & 0.4(0.9) & 1.0(1.2) & 1.1(1.2) & 0.7(0.5) & 0.7(0.5) & 0.8(1.3) & 0.7(1.3)  & 1.1(1.2)  & 1.1(1.3) \\[0pt]
&$0.2$ &0.5(0.4)& 0.6(0.4)  & 1.0(1.3)  & 0.9(1.2) & 0.9(1.0) & 1.0(1.1) & 0.8(0.6) & 0.9(0.6) & 0.8(1.1) & 0.8(1.1)  & 1.1(0.8)  & 1.3(1.1) \\[0pt]
&$0.4$ &0.7(0.5)& 1.7(0.8)  & 1.1(1.3)  & 2.5(2.3) & 0.9(0.7) & 2.7(1.5) & 0.7(0.5) & 1.9(0.8) & 1.0(1.2) & 1.9(1.9)  & 1.1(1.0)  & 2.7(2.0) \\[0pt]
&$0.6$ &0.7(0.5)& 1.8(0.9)  & 0.9(1.1)  & 2.8(2.6) & 1.1(0.8) & 2.7(1.9) & 0.6(0.4) & 1.9(0.8) & 0.8(1.3) & 3.0(2.6)  & 1.1(0.9)  & 2.8(1.7) \\[0pt]
\hline
\multirow{5}{*}{$(200,200)$} 
&$0.0$ &0.7(0.5)& 0.7(0.5)  & 0.8(0.9)  & 0.8(0.9) & 0.9(0.4) & 0.9(0.4) & 0.8(0.6) & 0.8(0.5) & 0.9(1.0) & 0.8(0.9)  & 0.8(0.4)  & 0.8(0.4) \\[0pt]
&$0.2$ &0.9(0.5)& 1.0(0.5)  & 1.1(1.1)  & 1.1(1.1) & 0.7(0.4) & 0.8(0.4) & 0.8(0.6) & 0.8(0.6) & 1.1(1.2) & 1.0(1.0)  & 0.7(0.4)  & 0.8(0.4) \\[0pt]
&$0.4$ &1.0(0.7)& 2.2(0.9)  & 1.0(1.0)  & 2.6(1.9) & 0.9(0.5) & 1.8(0.7) & 0.9(0.5) & 2.0(0.9) & 1.1(1.2) & 2.2(1.8)  & 0.8(0.4)  & 1.8(0.6) \\[0pt]
&$0.6$ &0.7(0.5)& 2.3(0.8)  & 1.1(0.9)  & 2.9(1.6) & 0.8(0.4) & 1.8(0.6) & 0.8(0.5) & 2.1(0.7) & 1.0(1.0) & 3.2(1.6)  & 0.8(0.3)  & 1.9(0.6) \\[0pt]
\hline
\multirow{5}{*}{$(800,200)$} 
&$0.0$ &0.8(0.3)& 0.8(0.3)  & 0.9(0.7)  & 0.9(0.7) & 0.7(0.2) & 0.7(0.2) & 0.7(0.2) & 0.7(0.2) & 0.9(0.6) & 0.9(0.6)  & 0.8(0.2)  & 0.8(0.2) \\[0pt]
&$0.2$ &0.8(0.3)& 0.9(0.3)  & 0.9(0.6)  & 0.9(0.7) & 0.7(0.2) & 0.8(0.2) & 0.7(0.2) & 0.9(0.3) & 1.0(0.7) & 1.0(0.7)  & 0.7(0.2)  & 0.8(0.2) \\[0pt]
&$0.4$ &0.8(0.2)& 2.5(0.5)  & 0.9(0.6)  & 2.1(0.9) & 0.7(0.2) & 2.2(0.3) & 0.7(0.2) & 2.5(0.5) & 1.0(0.7) & 2.1(1.0)  & 0.8(0.2)  & 2.3(0.4) \\[0pt]
&$0.6$ &0.8(0.2)& 2.4(0.4)  & 0.9(0.6)  & 2.2(0.9) & 0.8(0.2) & 2.3(0.4) & 0.8(0.2) & 2.4(0.4) & 0.9(0.5) & 2.2(0.9)  & 0.8(0.2)  & 2.3(0.4) \\[0pt]
\hline
& &\multicolumn{12}{c}{Empirical power (in \%)}\\[0pt]
\hline
\multirow{5}{*}{$(200,50)$}  
&$0.0$ &91.0& 90.9  & 52.8  & 53.4 & 18.3 & 18.2 & 91.3 & 91.3 & 54.8 & 54.4  & 19.1  & 19.2 \\[0pt]
&$0.2$ &88.1& 89.3  & 53.3  & 52.9 & 20.5 & 19.7 & 90.2 & 90.9 & 51.6 & 50.7  & 19.6  & 18.8 \\[0pt]
&$0.4$ &84.9& 89.9  & 58.9  & 56.4 & 25.3 & 20.7 & 86.2 & 90.8 & 57.1 & 53.0  & 23.3  & 18.3 \\[0pt]
&$0.6$ &85.6& 90.9  & 60.3  & 54.3 & 24.8 & 19.9 & 85.3 & 90.4 & 59.7 & 54.8  & 25.4  & 20.3 \\[0pt]     
\hline
\multirow{5}{*}{$(200,200)$}
&$0.0$ &100.0& 100.0  & 100.0  & 100.0 & 99.9 & 99.9 & 100.0 & 100.0 & 100.0 & 100.0  & 99.9  & 99.9 \\[0pt]
&$0.2$ &100.0& 100.0  & 100.0  & 100.0 & 99.9 & 99.9 & 100.0 & 100.0 & 100.0 & 100.0  & 99.9  & 99.9 \\[0pt]
&$0.4$ &100.0& 100.0  & 99.9  & 100.0 & 99.6 & 99.9 & 100.0 & 100.0 & 100.0 & 100.0  & 99.4  & 99.8 \\[0pt]
&$0.6$ &100.0& 100.0  & 99.9  & 100.0 & 99.5 & 99.9 & 100.0 & 100.0 & 100.0 & 100.0  & 99.7  & 99.9 \\[0pt]     
\hline
\multirow{5}{*}{$(800,200)$}
&$0.0$ &100.0& 100.0  & 61.1  & 60.6 & 99.6 & 99.6 & 100.0 & 100.0 & 61.5 & 60.9  & 99.6  & 99.6 \\[0pt]
&$0.2$ &100.0& 100.0  & 61.8  & 61.8 & 99.5 & 99.6 & 100.0 & 100.0 & 62.4 & 62.6  & 99.6  & 99.6 \\[0pt]
&$0.4$ &100.0& 100.0  & 64.5  & 63.3 & 98.5 & 99.5 & 100.0 & 100.0 & 64.1 & 63.0  & 98.6  & 99.5 \\[0pt]
&$0.6$ &100.0& 100.0  & 65.1  & 62.9 & 98.6 & 99.5 & 100.0 & 100.0 & 64.4 & 62.8  & 98.6  & 99.5 \\[0pt]     
\hline
\end{tabular}
\end{lrbox}
\scalebox{1.0}{\usebox{\tablebox}}
\label{tbl:general_down}
\end{center}
\end{sidewaystable}

\begin{sidewaystable}[hptb]\small\addtolength{\tabcolsep}{-4pt}
\begin{center}
\caption{Sensitivity I (top): The empirical FDR (with standard error in the parenthesis) and the empirical power, both in percentage, for the sensitivity analysis on the normality. Sensitivity II (bottom): The empirical FDR (with standard error in the parenthesis) and the empirical power, both in percentage, for the sensitivity analysis on the Kronecker structure of the off-diagonal block, i.e., $\Cov (\text{vec}\{\bold X^{(1)}\}, \text{vec}\{\bold X^{(2)}\}) = \S_{S_{1,2}} \otimes \S_{T_{1,2}}$.}
\label{tbl:sensitivity}
    \begin{lrbox}{\tablebox}
      \begin{tabular}{c@{\hspace{1em}} |c@{\hspace{1em}}  |r@{\hspace{1em}} r@{\hspace{1em}}|r@{\hspace{1em}} r @{\hspace{1em}}|r@{\hspace{1em}}r@{\hspace{1em}} |r@{\hspace{1em}} r@{\hspace{1em}}|r@{\hspace{1em}} r @{\hspace{1em}}|r@{\hspace{1em}}r@{\hspace{1em}} }
    \hline
    \multicolumn{14}{c}{\textbf{Sensitivity I}}\\[0pt]        
    \hline
    \multicolumn{2}{c|}{temporal structure}&\multicolumn{6}{c}{moving average}&\multicolumn{6}{|c}{autoregressive}\\[0pt]
    \hline
    \multicolumn{2}{c|}{spatial structure} &\multicolumn{2}{c}{banded}&\multicolumn{2}{|c}{hub}&\multicolumn{2}{|c}{small}&\multicolumn{2}{|c}{banded}&\multicolumn{2}{|c}{hub}&\multicolumn{2}{|c}{small}\\  [0pt]
    \hline
    \multicolumn{2}{c|}{correction}&\multicolumn{1}{c|}{\cmark}&\multicolumn{1}{c|}{\xmark}&\multicolumn{1}{c|}{\cmark}&\multicolumn{1}{c|}{\xmark}&\multicolumn{1}{c|}{\cmark}&\multicolumn{1}{c|}{\xmark}&\multicolumn{1}{c|}{\cmark}&\multicolumn{1}{c|}{\xmark}&\multicolumn{1}{c|}{\cmark}&\multicolumn{1}{c|}{\xmark}&\multicolumn{1}{c|}{\cmark}&\multicolumn{1}{c|}{\xmark}\\[0pt]
    \hline\hline\hline
    $(p,q)$  & $df$ &\multicolumn{12}{c}{Empirical FDR (SE)  (in \%)}\\[0pt]
    \hline
    \multirow{5}{*}{$(200,50)$}  
    &$4$ &0.7(0.5)& 1.9(0.8)  & 1.1(1.6)  & 4.1(2.6) & 1.1(0.8) & 2.7(1.4) & 0.8(0.6) & 1.8(0.9) & 1.2(1.5) & 4.3(3.2)  & 1.2(0.8)  & 2.8(1.5) \\[0pt]
    &$6$ &0.8(0.4)& 1.8(0.8)  & 1.0(1.3)  & 3.7(2.6) & 1.4(0.9) & 2.8(1.4) & 0.6(0.4) & 1.7(0.8) & 0.9(1.2) & 3.9(2.3)  & 1.3(1.0)  & 3.2(1.7) \\[0pt]
    &$8$ &0.7(0.5)& 1.6(0.8)  & 1.4(1.8)  & 3.6(2.6) & 1.3(1.1) & 2.6(1.3) & 0.6(0.5) & 1.8(0.8) & 0.6(1.1) & 3.2(2.7)  & 1.0(0.9)  & 2.9(1.5) \\[0pt]
    \hline
    \multirow{5}{*}{$(200,200)$} 
    &$4$ &0.8(0.6)& 1.9(0.8)  & 1.1(1.0)  & 3.3(1.8) & 0.9(0.4) & 2.1(0.6) & 0.7(0.5) & 2.0(0.8) & 0.9(1.2) & 3.2(2.0)  & 0.9(0.4)  & 1.9(0.6) \\[0pt]
    &$6$ &0.8(0.5)& 2.0(0.9)  & 0.8(0.8)  & 3.1(1.9) & 0.9(0.4) & 1.8(0.6) & 0.9(0.6) & 2.1(0.9) & 0.9(1.0) & 3.1(2.0)  & 0.8(0.5)  & 1.7(0.6) \\[0pt]
    &$8$ &1.0(0.5)& 2.1(0.9)  & 0.7(0.9)  & 3.2(1.9) & 0.8(0.4) & 1.9(0.6) & 0.8(0.6) & 2.1(0.7) & 1.0(1.1) & 3.2(1.6)  & 0.8(0.4)  & 1.9(0.6) \\[0pt]    
    \hline
    \multirow{5}{*}{$(800,200)$} 
    &$4$ &0.8(0.3)& 2.6(0.4)  & 0.9(0.7)  & 2.6(1.1) & 0.8(0.2) & 2.2(0.3) & 0.8(0.3) & 2.7(0.5) & 1.0(0.5) & 2.4(1.1)  & 0.8(0.2)  & 2.2(0.3) \\[0pt]
    &$6$ &0.8(0.3)& 2.5(0.4)  & 0.9(0.6)  & 2.4(0.8) & 0.8(0.2) & 2.1(0.3) & 0.8(0.3) & 2.5(0.5) & 1.0(0.6) & 2.3(0.9)  & 0.8(0.2)  & 2.1(0.3) \\[0pt]
    &$8$ &0.8(0.3)& 2.4(0.4)  & 0.8(0.6)  & 2.2(1.1) & 0.8(0.2) & 2.1(0.3) & 0.8(0.2) & 2.5(0.4) & 0.8(0.5) & 2.0(0.7)  & 0.8(0.2)  & 2.2(0.3) \\[0pt]    
    \hline
    & &\multicolumn{12}{c}{Empirical power (SE)  (in \%)}\\[0pt]
    \hline
    \multirow{5}{*}{$(200,50)$}  
    &$4$ &90.3& 94.4  & 59.4  & 52.5 & 29.5 & 23.6 & 90.4 & 94.5 & 58.7 & 53.4  & 28.9  & 23.3 \\[0pt]
    &$6$ &90.3& 94.5  & 59.2  & 53.4 & 28.5 & 23.2 & 90.9 & 94.9 & 58.0 & 51.3  & 27.8  & 21.9 \\[0pt]
    &$8$ &90.7& 94.8  & 59.1  & 52.7 & 29.3 & 22.7 & 90.6 & 95.1 & 59.7 & 53.6  & 28.7  & 23.0 \\[0pt]    
    \hline
    \multirow{5}{*}{$(200,200)$}
    &$4$ &100.0& 100.0  & 99.9  & 100.0 & 99.7 & 99.9 & 100.0 & 100.0 & 99.9 & 100.0  & 99.6  & 99.9 \\[0pt]
    &$6$ &100.0& 100.0  & 99.9  & 100.0 & 99.7 & 99.9 & 100.0 & 100.0 & 99.9 & 100.0  & 99.7  & 99.9 \\[0pt]
    &$8$ &100.0& 100.0  & 99.9  & 100.0 & 99.8 & 99.9 & 100.0 & 100.0 & 99.9 & 100.0  & 99.7  & 99.9 \\[0pt]    
    \hline
    \multirow{5}{*}{$(800,200)$}
    &$4$ &100.0& 100.0  & 64.6  & 66.4 & 98.9 & 99.6 & 100.0 & 100.0 & 65.3 & 67.2  & 98.9  & 99.6 \\[0pt]
    &$6$ &100.0& 100.0  & 65.1  & 66.6 & 99.0 & 99.6 & 100.0 & 100.0 & 65.3 & 67.4  & 98.9  & 99.6 \\[0pt]
    &$6$ &100.0& 100.0  & 64.6  & 66.4 & 99.0 & 99.7 & 100.0 & 100.0 & 65.5 & 66.5  & 99.0  & 99.6 \\[0pt]    
      \hline
    \end{tabular}
  \end{lrbox}  
    \scalebox{0.8}{\usebox{\tablebox}}

        \begin{lrbox}{\tablebox}
          \begin{tabular}{c@{\hspace{1em}} |c@{\hspace{1em}}  |r@{\hspace{1em}} r@{\hspace{1em}}|r@{\hspace{1em}} r @{\hspace{1em}}|r@{\hspace{1em}}r@{\hspace{1em}} |r@{\hspace{1em}} r@{\hspace{1em}}|r@{\hspace{1em}} r @{\hspace{1em}}|r@{\hspace{1em}}r@{\hspace{1em}} }
        \multicolumn{14}{c}{\textbf{Sensitivity II}}\\[0pt]        
    \hline
    \multicolumn{2}{c|}{temporal structure}&\multicolumn{6}{c}{moving average}&\multicolumn{6}{|c}{autoregressive}\\[0pt]
    \hline
    \multicolumn{2}{c|}{spatial structure} &\multicolumn{2}{c}{banded}&\multicolumn{2}{|c}{hub}&\multicolumn{2}{|c}{small}&\multicolumn{2}{|c}{banded}&\multicolumn{2}{|c}{hub}&\multicolumn{2}{|c}{small}\\  [0pt]
    \hline
    \multicolumn{2}{c|}{correction}&\multicolumn{1}{c|}{\cmark}&\multicolumn{1}{c|}{\xmark}&\multicolumn{1}{c|}{\cmark}&\multicolumn{1}{c|}{\xmark}&\multicolumn{1}{c|}{\cmark}&\multicolumn{1}{c|}{\xmark}&\multicolumn{1}{c|}{\cmark}&\multicolumn{1}{c|}{\xmark}&\multicolumn{1}{c|}{\cmark}&\multicolumn{1}{c|}{\xmark}&\multicolumn{1}{c|}{\cmark}&\multicolumn{1}{c|}{\xmark}\\[0pt]
    \hline\hline\hline
    $p^{\star}$  & $l^{\star}$ &\multicolumn{12}{c}{Empirical FDR (SE)  (in \%)}\\[0pt]
    \hline
      \multirow{2}{*}{$0$}
    &$0.1$ &0.7(0.4)& 0.7(0.5)  & 0.7(1.2)  & 1.1(1.6) & 1.1(0.9) & 1.2(1.0) & 0.7(0.5) & 0.7(0.5) & 0.8(1.1) & 1.3(1.4)  & 1.0(1.1)  & 1.1(1.0) \\[0pt]
    &$1$ &0.6(0.5)& 0.7(0.6)  & 1.2(1.9)  & 1.5(1.9) & 1.3(1.1) & 1.4(1.0) & 0.6(0.6) & 0.7(0.5) & 0.9(1.3) & 1.2(1.5)  & 1.3(0.9)  & 1.4(1.2) \\[0pt]
    \hline
    \multirow{2}{*}{$1$}  
    &$0.1$ &0.7(0.5)& 0.7(0.5)  & 1.0(1.6)  & 1.1(1.7) & 0.9(0.9) & 1.0(0.9) & 0.7(0.4) & 0.7(0.4) & 0.7(1.1) & 1.1(1.3)  & 1.3(0.9)  & 1.3(1.0) \\[0pt]
    &$1$ &0.8(0.5)& 0.8(0.5)  & 0.7(1.2)  & 0.8(1.2) & 1.2(1.0) & 1.2(1.1) & 0.7(0.5) & 0.7(0.4) & 0.9(1.2) & 0.9(1.1)  & 1.1(0.9)  & 1.1(0.9) \\[0pt]
    \hline
    \multirow{2}{*}{$5$} 
    &$0.1$ &0.6(0.5)& 0.7(0.5)  & 0.9(1.3)  & 1.4(1.5) & 1.0(0.9) & 1.2(1.0) & 0.7(0.5) & 0.7(0.5) & 0.7(1.2) & 0.7(1.4)  & 1.2(1.0)  & 1.3(1.1) \\[0pt]
    &$1$ &0.8(0.6)& 0.8(0.6)  & 1.3(1.4)  & 1.4(1.4) & 1.0(0.8) & 1.1(0.9) & 0.7(0.4) & 0.7(0.5) & 0.5(1.0) & 0.6(1.3)  & 0.9(0.9)  & 1.1(1.0) \\[0pt]
    \hline
    \multirow{2}{*}{$10$} 
    &$0.1$ &0.9(0.6)& 0.9(0.5)  & 0.9(1.2)  & 0.8(1.1) & 1.2(1.0) & 1.3(1.1) & 0.9(0.7) & 0.9(0.6) & 1.0(1.3) & 1.1(1.6)  & 1.1(0.8)  & 1.1(0.8) \\[0pt]
    &$1$ &0.5(0.5)& 0.6(0.5)  & 1.1(1.7)  & 1.1(1.7) & 1.2(1.1) & 1.2(1.2) & 0.6(0.5) & 0.6(0.5) & 1.2(1.6) & 1.2(1.7)  & 1.1(0.9)  & 1.2(1.1) \\[0pt]
    \hline
    & &\multicolumn{12}{c}{Empirical power (SE)  (in \%)}\\[0pt]
    \hline
    \multirow{2}{*}{$0$}  
    &$0.1$ &94.6& 95.0  & 50.9  & 49.3 & 23.8 & 23.5 & 94.7 & 95.2 & 53.4 & 52.2  & 23.2  & 22.7 \\[0pt]
    &$1$ &95.1& 95.2  & 54.0  & 52.0 & 22.9 & 22.6 & 95.1 & 95.4 & 51.9 & 51.0  & 23.2  & 22.9 \\[0pt]
    \hline
    \multirow{2}{*}{$1$}  
    &$0.1$ &94.8& 95.0  & 53.8  & 52.8 & 23.2 & 22.9 & 94.8 & 95.1 & 52.9 & 51.8  & 23.3  & 23.0 \\[0pt]
    &$1$ &94.6& 94.9  & 53.6  & 53.1 & 23.2 & 22.6 & 95.2 & 95.4 & 51.8 & 50.9  & 23.0  & 22.5 \\[0pt]
    \hline
    \multirow{2}{*}{$5$}
    &$0.1$ &94.8& 95.2  & 52.1  & 52.1 & 23.3 & 22.9 & 94.7 & 95.0 & 53.3 & 51.9  & 23.5  & 23.1 \\[0pt]
    &$1$ &94.9& 95.2  & 52.2  & 52.0 & 22.9 & 22.6 & 94.6 & 95.0 & 52.1 & 52.0  & 23.1  & 23.1 \\[0pt]
    \hline
    \multirow{2}{*}{$10$}
    &$0.1$ &94.6& 95.1  & 53.3  & 51.6 & 23.3 & 22.8 & 94.7 & 94.8 & 52.7 & 51.5  & 22.9  & 22.3 \\[0pt]
    &$1$ &95.1& 95.4  & 52.8  & 52.1 & 22.8 & 22.7 & 94.7 & 94.9 & 52.6 & 51.6  & 23.8  & 23.5 \\[0pt]
    \hline
    \end{tabular}
    \end{lrbox}
    \scalebox{0.8}{\usebox{\tablebox}}

\end{center}
\end{sidewaystable}

\begin{table}[t!]
  \caption{Top 30 differentiating links of the brain connectivity networks of the 23 subjects of the ADNI database before and after the conversion from MCI to AD. The last column shows the direction of the link change. ``+'' represents the link gets enhanced after the conversion, and ``-'' represents the link gets weakened after the conversion.}
\label{tbl:top30_real}
\begin{center}
\begin{tabular}{lllc}
\hline\hline
rank & Differentiating links               & $p$-value              & +/- \\ \hline 
1&Cerebellum\_Crus1\_L$\leftrightarrow$Temporal\_Inf\_R&0&-\\
2&Temporal\_Pole\_Sup\_R$\leftrightarrow$Occipital\_Mid\_L&1.11e-16&-\\
3&Temporal\_Pole\_Mid\_R$\leftrightarrow$Occipital\_Sup\_L&2.22-16&+\\
4&Temporal\_Pole\_Mid\_R$\leftrightarrow$Occipital\_Mid\_L&3.33-16&+\\
5&Paracentral\_Lobule\_R$\leftrightarrow$Rolandic\_Oper\_L&7.77e-16&+\\
6&Cerebellum\_Crus2\_L$\leftrightarrow$Frontal\_Sup\_Orb\_R&9.99e-15&-\\
7&Cerebellum\_7b\_L$\leftrightarrow$Frontal\_Sup\_Orb\_R&1.07e-14&-\\
8&Cerebellum\_7b\_R$\leftrightarrow$Occipital\_Mid\_R&1.74e-14&+\\
9&Cerebellum\_8\_R$\leftrightarrow$Calcarine\_R&2.72e-14&+\\
10&Temporal\_Inf\_L$\leftrightarrow$Fusiform\_L&3.30e-14&-\\
11&Fusiform\_R$\leftrightarrow$Cuneus\_R&2.02e-13&-\\
12&Cerebellum\_Crus2\_L$\leftrightarrow$Temporal\_Inf\_R&3.85e-13&-\\
13&Occipital\_Inf\_R$\leftrightarrow$Rectus\_L&7.37e-13&+\\
14&Cerebellum\_7b\_L$\leftrightarrow$Fusiform\_R&9.01e-13&+\\
15&ParaHippocampal\_R$\leftrightarrow$Frontal\_Inf\_Orb\_L&1.62e-12&+\\
16&Temporal\_Pole\_Mid\_R$\leftrightarrow$Temporal\_Pole\_Sup\_L&2.77e-12&+\\
17&Heschl\_L$\leftrightarrow$Lingual\_R&3.22e-12&-\\
18&Cerebellum\_10\_R$\leftrightarrow$Olfactory\_R&3.46e-12&+\\
19&Cerebellum\_9\_L$\leftrightarrow$Frontal\_Mid\_Orb\_L&4.89e-12&-\\
20&Cerebellum\_Crus1\_R$\leftrightarrow$Cerebellum\_Crus1\_L&9.00e-12&-\\
21&Cerebellum\_10\_R$\leftrightarrow$Frontal\_Mid\_Orb\_R&1.42e-11&+\\
22&Cerebellum\_10\_L$\leftrightarrow$Frontal\_Mid\_Orb\_R&1.75e-11&-\\
23&Cerebellum\_Crus1\_L$\leftrightarrow$Frontal\_Mid\_Orb\_L&1.96e-11&+\\
24&SupraMarginal\_L$\leftrightarrow$Cuneus\_R&2.45e-11&-\\
25&Cerebellum\_6\_L$\leftrightarrow$Cerebellum\_Crus2\_L&3.70e-11&+\\
26&Cerebellum\_7b\_L$\leftrightarrow$Rectus\_L&4.96e-11&-\\
27&Cerebellum\_3\_R$\leftrightarrow$Frontal\_Med\_Orb\_R&5.69e-11&+\\
28&Insula\_R$\leftrightarrow$Frontal\_Inf\_Oper\_R&5.88e-11&-\\
29&Angular\_R$\leftrightarrow$Angular\_L&7.00e-11&-\\
30&Cerebellum\_7b\_L$\leftrightarrow$Temporal\_Inf\_R&7.02e-11&+\\
\hline
\end{tabular}
\end{center}
\end{table}

\clearpage
\appendix
\begin{center}
  {\huge Appendix}
\end{center}

\section{Regularity conditions}
\label{sec:theory:conditions}

We first introduce a set of regularity conditions that are required to establish the asymptotic properties for our proposed testing procedure. In the following, we note that, Condition (A5) is for the case when $\S_{T_t}$ and $\S_{T_{1,2}}$ are known, and (A6)-(A7) are for the case when $\S_{T_t}$ and $\S_{T_{1,2}}$ are unknown. Conditions (A1)-(A4) are required for both cases. Denote $\lambda_{\min}(\cdot)$ and $\lambda_{\max}(\cdot)$ as the smallest and the largest eigenvalue, respectively, and $\sigma_{\max}(\cdot)$ as the largest singular value.

\begin{enumerate}[({A}1)]
\item There are constants $c_0, c_1, c_2 > 0$ such that (i) $c_0^{-1}\leq \underset{t = 1, 2}{\min}\{\lambda_{\min}(\O_{T_t}), \lambda_{\min}(\O_{S_t})\}\leq \underset{t = 1, 2}{\max}\{\lambda_{\max}(\O_{T_t}), \lambda_{\max}(\O_{S_t})\}\leq c_0$; (ii) $\max\{\sigma_{\max}(\Rho_{T_{1,2}})|\text{tr}(\S_{S_{1,2}})|/p, \; \sigma_{\max}(\S_{S_{1,2}})|\text{tr}(\Rho_{T_{1,2}})|$ $/q\}\leq c_1$; and (iii) $|\text{tr}(\Rho_{T_{1,2}})\text{tr}(\S_{S_{1,2}})|/(pq)\geq c_2$. 

\item Let $A_{\tau} = \{(i, j): |\O_{S_t, i, j}| \geq (\log p)^{-2-\tau},1\leq i< j \leq p, t=1,2\}$. There exists some $\tau > 0$ such that $|A_{\tau} \cap \mathcal{H}_0| = o(p^{\nu})$ for any $\nu > 0$. 

\item Assume that $\log p = o\{(nq)^{1/5}\}$, and $q  = o\{np/(\log p)^2\}$.

\item Let $\mathcal{S}(\eta) = \left\{(i,j): 1 \leq i < j \leq p,  \frac{|\rho_{S_{1}, i, j} - \rho_{S_{2}, i, j}|}{\Theta_{i,j}^{1/2}} \geq (\log p)^{1/2 + \eta}\right\}$. For some $\eta$, $\delta > 0$, $|\mathcal{S}(\eta)| \geq [1/\{(8\pi)^{1/2} \alpha\} + \delta](\log \log p)^{1/2}$.

\item When $\S_{T_t}$ and $\S_{T_{1,2}}$ are known, denote the corresponding regression coefficient estimate by $\hat\be_{i}^{(t)}$. Assume that $\underset{1\leq i \leq p}{\max} \|\hat\be_{i}^{(t)} - \be_{i}^{(t)}\|_1 = o_p\left[ \{\log \max (p, q, n)\}^{-1} \right]$, and $\underset{1 \leq i \leq p}{\max} \|\hat\be_{i}^{(t)} - \be_{i}^{(t)}\|_2 = o_p\left\{ (nq\log p)^{-1/4} \right\}$. 

\item When $\S_{T_t}$ and $\S_{T_{1,2}}$ are unknown, denote the corresponding regression coefficient estimate by $\hat{\be}_{i}^{(t,d)}$. Assume that $\underset{1\leq i \leq p, 1 \leq l \leq q}{\max} \|\hat{\be}_{i}^{(t,d)} - \be_{i,l}^{(t)}\|_1 =  o_p\left[ \{\log \max (p, q, n)\}^{-1} \right]$, and $\underset{1\leq i \leq p, 1 \leq l \leq q}{\max} \|\hat{\be}_{i}^{(t,d)} - \be_{i,l}^{(t)}\|_2 = o_p\left\{ (nq\log p)^{-1/4} \right\}$.

\item When $\S_{T_t}$ and $\S_{T_{1,2 }}$ are unknown, denote the estimator of $\S_{T_t}$ as $\hat\S_{T_t}$. Assume $\|\hat{\S}_{T_t}^{-1/2} - c \S_{T_t}^{-1/2}\|_{\max} = o_p(r_{n,p,q})$ for some constant $c$, where $r_{n,p,q} = o\{\min_t \min (r_{n, p, q, t}^{(1)}, r_{n, p, q, t}^{(1)})\}$, $r_{n, p, q, t}^{(1)} = [s_{p,q} q \{ \log \max (p, q, n)^{3/2} \|\O_{S_t}\|_{L_1}^2 \} ]^{-1}$, $r_{n, p, q, t}^{(2)} = [nqs_{p,q}^2 \log p \{\log \max (p, q, n)\}^2]^{-1/4}$ $(q \|\O_{S_t}\|_{L_1}^2 )^{-1}$, $s_{p,q} = \max_{t \in \{1,2\}} \max_{1\leq l \leq q} \max_{1 \leq i \leq p} \sum_{j = 1}^p \max\{\mathbb{I}(\omega_{S_t, i, j} \neq 0), \mathbf{1}(\hat{\omega}_{l, i, j}^{(t,d)} \neq 0)\}$, $(\hat{\omega}_{l,i,j}^{(t,d)})_{p\times p} = \left \{\Cov \left( \X_k^{(t)} \hat{\S}_{T_t}^{-1/2} \right) \right \}^{-1}$, and $\mathbf{1}(\cdot)$ is the indicator function. Here, for a matrix $\A \in \real{p\times p}$, $\|\A\|_{\max} = \underset{1 \leq i, j \leq p}{\max} |a_{i,j}|$ is the matrix element-wise max norm, and $\|\A\|_{L_1} = \underset{1 \leq j \leq p}{\max} \sum_{i = 1}^p |a_{i,j}|$ is the matrix 1-norm. 
\end{enumerate}

Most of the above conditions are parallel to those for the two-sample test of matrix graphs of \citet{xia2018two}. Here we make remarks on a few different ones. In (A1), we have added two conditions, (ii) $\max\{\sigma_{\max}(\Rho_{T_{1,2}})|\text{tr}(\S_{S_{1,2}})|/p,\sigma_{\max}(\S_{S_{1,2}})|\text{tr}(\Rho_{T_{1,2}})|/q\}\leq c_1$, and (iii) $|\text{tr}(\Rho_{T_{1,2}})$ $\text{tr}(\S_{S_{1,2}})|/(pq) \geq c_2$. These are purely technical assumptions to simplify the proofs, and we view both conditions mild. Note that the three terms, $\sigma_{\max}(\Rho_{T_{1,2}})|\text{tr}(\S_{S_{1,2}})|/p$, $\sigma_{\max}(\S_{S_{1,2}})|\text{tr}(\Rho_{T_{1,2}})|/q$,  and $|\text{tr}(\Rho_{T_{1,2}})\text{tr}(\S_{S_{1,2}})|/(pq)$, are all identifiable. It can be easily shown that both singular values, $\sigma_{\max}(\Rho_{T_{1,2}})|\text{tr}(\S_{S_{1,2}})|/p$ and $\sigma_{\max}(\S_{S_{1,2}})|\text{tr}(\Rho_{T_{1,2}})|/q$, are bounded if $\S$ defined in \eqref{eq:joint_dist_X} of the paper has bounded eigenvalues. Furthermore, the before and after stimulus observations at the same time points or spatial locations are likely to have non-trivial temporal or spatial  dependency, and thus it is reasonable to assume $|\text{tr}(\Rho_{T_{1,2}})\text{tr}(\S_{S_{1,2}})|/(pq)$ to be bounded away from zero. In (A3), we have added that $q  = o\{np/(\log p)^2\}$. This is to ensure the convergence rate under the Frobenius norm and the maximum norm for the estimate of $\Rho_{T_{1,2}}$ in computing the empirical covariance matrix by pooling $2p$ spatial locations. Again it is a mild technical condition.

\section{Technical lemmas}
\label{sec:theory:lemma}

We next introduce some technical lemmas that are useful for the subsequent proofs. Define $U_{i,j}^{(t)}=\frac{1}{nq}\sum_{k=1}^{n}\sum_{l=1}^q \Big\{ \epsilon^{(t)}_{k,i,l}\epsilon^{(t)}_{k,j,l}-\EE\epsilon^{(t)}_{k,i,l}\epsilon^{(t)}_{k,j,l} \Big\}$, $(\hat{\sigma}^{(t)}_{i,j,\epsilon})=\frac{1}{nq}\sum_{k=1}^{n}\sum_{l=1}^q(\eps^{(t)}_{k,\cdot,l}-\bar{\eps}^{(t)})(\eps^{(t)}_{k,\cdot,l}-\bar{\eps}^{(t)})'$, $\eps^{(t)}_{k,\cdot,l}=(\epsilon^{(t)}_{k,1,l},\dots,\epsilon^{(t)}_{k,p,l})$, and $\bar{\eps}^{(t)}=\frac{1}{nq}\sum_{k=1}^{n}\sum_{l = 1}^{q}\eps^{(t)}_{k,\cdot,l}$, $t=1,2$. 

\begin{lemma}\label{lemma:hatrij_oracle}
Suppose that (A1), (A3) and (A5) hold. Then we have
\begin{eqnarray*}
& & \max_{1\leq i\leq p}|\hat{r}_{i,i}^{(t)}-r_{i,i}^{(t)}|=O_{\rm p}[\{{\log p/(nq)}\}^{1/2}], \\
& & \tilde{r}_{i,j}^{(t)}=\tilde{R}_{i,j}^{(t)}-\tilde{r}_{i,i}^{(t)}(\hat{\beta}_{i,j}^{(t)}-\beta_{i,j}^{(t)})-\tilde{r}_{j,j}^{(t)}(\hat{\beta}_{j-1,i}^{(t)}-\beta_{j-1,i}^{(t)})+o_{\rm p}\{(nq\log p)^{-1/2}\},
\end{eqnarray*}
  for $1\leq i<j\leq p$, $t=1,2$, where $\tilde{R}_{i,j}^{(t)}$ is the empirical covariance between $\{\epsilon^{(t)}_{k,i,l}, k=1,\dots,n, l=1,\dots,q\}$ and $\{\epsilon^{(t)}_{k,j,l}, k = 1, \dots, n, l=1,\dots,q\}$. Consequently, uniformly in $1\leq i<j\leq p$,
  \[
    \hat{r}_{i,j}^{(t)}-(\omega_{S_t,i,i}\hat{\sigma}^{(t)}_{i,i,\epsilon}+\omega_{S_t,j,j}\hat{\sigma}^{(t)}_{j,j,\epsilon}-1)r_{i,j}^{(t)}=-U_{i,j}^{(t)}+o_{\rm p}\{(nq\log p)^{-1/2}\}.
  \]
\end{lemma}
\noindent A similar lemma was proved in \citet{xia2015testing}, but we deal with $nq$ inverse regression models here.

\begin{lemma}\label{lemma:gaussian_ldb}
Let $\X_k\sim N(\m_1,\S_1)$, $k=1,\ldots,n_1$, and $\Y_k\sim N(\m_2,\S_2)$, $k=1,\ldots,n_2$. Define
\begin{eqnarray*}
    \tilde{\S}_{1}=(\tilde{\sigma}_{i,j}^{(1)})_{p\times p}=\frac{1}{n_{1}}\sum_{k=1}^{n_{1}}(\X-\m_{1})(\X-\m_{1})^{\T}, \quad \tilde{\S}_{2}=(\tilde{\sigma}_{i,j}^{(2)})_{p\times p}=\frac{1}{n_{2}}\sum_{k=1}^{n_{2}}(\Y-\m_{2})(\Y-\m_{2})^{\T}.
\end{eqnarray*}
Then, for some constant $C>0$, $\tilde{\sigma}_{i,j}^{(1)}-\tilde{\sigma}_{i,j}^{(2)}$ satisfies the  large deviation bound,
\begin{eqnarray*}
    &&\P\left[ \max_{(i,j)\in \mathcal{S}}\frac{\left\{ \tilde{\sigma}_{i,j}^{(1)}-\tilde{\sigma}_{i,j}^{(2)}-\sigma_{i,j}^{(1)}+\sigma_{i,j}^{(2)} \right\}^{2}}{\Var\{(X_{k,i}-\mu_{1,i})(X_{k,j}-\mu_{1,j})\}/n_{1}+
       \Var\{(Y_{k,i}-\mu_{2,i})(Y_{k,j}-\mu_{2,j})\}/n_{2}}\geq x^{2} \right] \cr
       \quad && \leq C|\mathcal{S}|\{1-\Phi(x)\}+O(p^{-1}), 
\end{eqnarray*}
uniformly for $0\leq x\leq (8\log p)^{1/2}$ and any subset $\mathcal{S}\subseteq \{(i,j):1\leq i\leq j\leq p\}$.
\end{lemma}
\noindent This lemma was proved in Lemma 4 of \citet{cai2013two}.

\begin{lemma}\label{lemma:hatrij_data}
Suppose (A1), (A3), (A6) and (A7) hold, then uniformly in $1\leq i\leq j\leq p$,
  \begin{equation*}
  \hat{r}_{i,j}^{(t,d)}-(\omega_{S_t,i,i}\hat{\sigma}^{(t)}_{i,i,\epsilon}+\omega_{S_t,j,j}\hat{\sigma}^{(t)}_{j,j,\epsilon}-1)r_{i,j}^{(t)}=-U_{i,j}^{(t)}+o_{\rm p}\{(nq\log p)^{-1/2}\},
  \end{equation*}
  for $t=1,2$, where $\hat{r}_{i,j}^{(t,d)}=-(\tilde{r}_{i,j}^{(t,d)}+\tilde{r}_{i,i}^{(t,d)}\hat{\beta}_{i,j}^{(t,d)}+\tilde{r}_{j,j}^{(t,d)}\hat{\beta}_{j-1,i}^{(t,d)})$, $\tilde{r}_{i,j}^{(t,d)}=1/(nq)\sum_{k=1}^n\sum_{l=1}^q\hat{\epsilon}_{k,i,l}^{(t,d)}\hat{\epsilon}_{k,j,l}^{(t,d)}$ and $\hat{\epsilon}_{k,i,l}^{(t,d)}=Y_{k,i,l}^{(t,d)} - \bar{Y}_{\cdot,i,\cdot}^{(t,d)}-\{(\Y_{k,-i,l}^{(t,d)})^{\T} - (\bar{\bold Y}_{\cdot,-i,\cdot}^{(t,d)})^{\T}\}\hat{\be}_{i,l}^{(t,d)}$ with $\Y_{k}^{(t,d)}=\X_k^{(t)}\hat{\S}_{T_t}^{-1/2}$.
\end{lemma}
\noindent This lemma was essentially proved in the proofs of Theorems 5 and 6 in \citet{xia2017one}.

\section{Proofs}

We now prove the propositions and theorem in the paper. Let $a_{n1}$  and $a_{n2}$ satisfy $\max_{1\leq i \leq p} \|\hat{\be}_{i}^{(t)} - \be_{i}^{(t)}\|_1 = O_{\rm p}(a_{n1})$ and $\max_{1\leq i \leq p} \|\hat{\be}_{i}^{(t)} - \be_{i}^{(t)}\|_2 = O_{\rm p}(a_{n2})$ when $\S_{T_t}$ and $\S_{T_{1,2}}$ are known, and satisfy $\max_{1\leq i \leq p, 1 \leq l \leq q} \|\hat{\be}_{i}^{(t,d)} - \be_{i,l}^{(t)}\|_1 = O_{\rm p}(a_{n1})$ and $\max_{1\leq i \leq p, 1 \leq l \leq q} \|\hat{\be}_{i}^{(t,d)} - \be_{i,l}^{(t)}\|_2 = O_{\rm p}(a_{n2})$ when $\S_{T_t}$ and $\S_{T_{1,2}}$ are unknown. Then $a_{n1} = o\left[ \{\log \max (p, q, n)\}^{-1} \right]$ and  $a_{n2} = o \left\{ (nq\log p)^{-1/4} \right\}$, respectively,  following Assumption (A5) or (A6).

\subsection{Proof of Proposition \ref{thm:var_correct}}
\noindent Note that
\begin{eqnarray}
  &&\Theta_{i,j} =  \Var(\tilde{U}_{i,j}^{(1)} - \tilde{U}_{i,j}^{(2)}) \nonumber \\
                                  &&\qquad = \Var(\tilde{U}_{i,j}^{(1)}) + \Var(\tilde{U}_{i,j}^{(2)}) - 2 \Cov(\tilde{U}_{i,j}^{(1)}, \tilde{U}_{i,j}^{(2)})\nonumber\\
                                  && \qquad = \theta_{i,j}^{(1)} + \theta_{i,j}^{(2)} - \frac{2/n^2q^2}{\sqrt{r_{i,i}^{(1)}r_{j,j}^{(1)}}\sqrt{r_{i,i}^{(2)}r_{j,j}^{(2)}}} \Cov(\sum_{k=1}^n\sum_{l=1}^q\epsilon_{k,i,l}^{(1)}\epsilon_{k,j,l}^{(1)}, \sum_{k=1}^n\sum_{l=1}^q\epsilon_{k,i,l}^{(2)}\epsilon_{k,j,l}^{(2)})\nonumber\\
                                  &&\qquad = \theta_{i,j}^{(1)} + \theta_{i,j}^{(2)} - \frac{2/nq^2}{\sqrt{r_{i,i}^{(1)}r_{j,j}^{(1)}}\sqrt{r_{i,i}^{(2)}r_{j,j}^{(2)}}} \sum_{l_1,l_2}^q \EE(\epsilon_{k,i,l_1}^{(1)}\epsilon_{k,j,l_1}^{(1)}\epsilon_{k,i,l_2}^{(2)}\epsilon_{k,j,l_2}^{(2)}) + \frac{2}{nq}  \frac{r_{i,j}^{(1)}}{\sqrt{r_{i,i}^{(1)}r_{j,j}^{(1)}}} \frac{r_{i,j}^{(2)}}{\sqrt{r_{i,i}^{(2)}r_{j,j}^{(2)}}}\nonumber 
\end{eqnarray}
By \eqref{eq:joint_dist_Y} and  \eqref{eq:inv_reg} in the paper, it follows that 
\begin{equation*}
  (\epsilon_{k,i,l_1}^{(1)}, \epsilon_{k,j,l_1}^{(1)}, \epsilon_{k,i,l_2}^{(2)}, \epsilon_{k,j,l_2}^{(2)})^T \sim N(0, \breve{\S}^{(i,j,l_1,l_2)}),
\end{equation*}
where $\breve{\S}^{(i,j,l_1,l_2)}$ is a positive semidefinitive matrix. Due to the fact that $\Var(\epsilon_{k, i, l}^{(t)}) = r_{i,i}^{(t)}$, we write 
\[ 
  \breve{\S}^{(i,j,l_1, l_2)} =
  \left({\begin{array}{cccc}
           r_{i,i}^{(1)} & \rho_{i,j}^{(1)}\sqrt{r_{i,i}^{(1)}r_{j,j}^{(1)}} &  \rho_{i,i;l_1,l_2}^{(1,2)}\sqrt{r_{i,i}^{(1)}r_{i,i}^{(2)}} &  \rho_{i,j;l_1,l_2}^{(1,2)}\sqrt{r_{i,i}^{(1)}r_{j,j}^{(2)}}\\
           \rho_{i,j}^{(1)}\sqrt{r_{i,i}^{(1)}r_{j,j}^{(1)}} & r_{j,j}^{(1)} &  \rho_{j,i;l_1,l_2}^{(1,2)}\sqrt{r_{j,j}^{(1)}r_{i,i}^{(2)}} &  \rho_{j,j;l_1,l_2}^{(1,2)}\sqrt{r_{j,j}^{(1)}r_{j,j}^{(2)}} \\
           \rho_{i,i;l_1,l_2}^{(1,2)}\sqrt{r_{i,i}^{(1)}r_{i,i}^{(2)}} &  \rho_{j,i;l_1,l_2}^{(1,2)}\sqrt{r_{j,j}^{(1)}r_{i,i}^{(2)}} & r_{i,i}^{(2)} &  \rho_{i,j}^{(2)}\sqrt{r_{i,i}^{(2)}r_{j,j}^{(2)}}\\
           \rho_{i,j;l_1,l_2}^{(1,2)}\sqrt{r_{i,i}^{(1)}r_{j,j}^{(2)}} &  \rho_{j,j;l_1,l_2}^{(1,2)}\sqrt{r_{j,j}^{(1)}r_{j,j}^{(2)}} &  \rho_{i,j}^{(2)}\sqrt{r_{i,i}^{(2)}r_{j,j}^{(2)}} &r_{j,j}^{(2)} \\
         \end{array}}
  \right),
\]
where $\rho_{i^{\prime},j^{\prime};l_1,l_2}^{(1,2)} = \text{corr} (\epsilon_{k,i^{\prime},l_1}^{(1)},\epsilon_{k,j^{\prime},l_2}^{(2)})$, $i^{\prime}, j^{\prime} \in \{i, j\}$; $\rho_{i,j}^{(t)} = \text{corr} (\epsilon_{k, i, l_1}^{(t)}, \epsilon_{k, j, l_2}^{(t)})$, for any $1 \leq l_1, l_2 \leq q$, $t = 1, 2$, and $\text{corr}(\cdot, \cdot)$ denotes the Pearson correlation between two random variables. It has been shown that $\rho_{i,j}^{(1)} = \frac{r_{i,j}^{(1)}}{\sqrt{r_{i,i}^{(1)}r_{j,j}^{(1)}}}$,  $\rho_{i,j}^{(2)} = \frac{r_{i,j}^{(2)}}{\sqrt{r_{i,i}^{(2)}r_{j,j}^{(2)}}}$. As for $\rho_{i,i;l_1,l_2}^{(1,2)}$, $\rho_{j,i;l_1,l_2}^{(1,2)}$
$\rho_{i,j;l_1,l_2}^{(1,2)}$ and $\rho_{j,j;l_1,l_2}^{(1,2)}$, it is easy to show that $\rho_{i^{\prime},j^{\prime};l_1,l_2}^{(1,2)} = \Rho_{T_{1,2}, l_1, l_2} \cdot \rho_{S_{1,2}, i^{\prime},j^{\prime}}$, where $\rho_{S_{1,2}, i^{\prime},j^{\prime}}:=\sqrt{r_{i^{\prime},i^{\prime}}^{(1)}r_{j^{\prime},j^{\prime}}^{(2)}} \cdot { \O_{S_1, i^{\prime}, \cdot}} \S_{S_{1,2}}\O_{S_2, \cdot, j^{\prime}}$ for $i^{\prime}, j^{\prime} \in \{i,j\}$.
By the property of an elliptically contoured distribution (\citealp{anderson2003}), it follows that
$$\EE(\epsilon_{k,i,l_1}^{(1)}\epsilon_{k,j,l_1}^{(1)}\epsilon_{k,i,l_2}^{(2)}\epsilon_{k,j,l_2}^{(2)}) = (\rho_{i,j}^{(1)}\rho_{i,j}^{(2)} + \rho_{i,i;l_1,l_2}^{(1,2)}\rho_{j,j;l_1,l_2}^{(1,2)} + \rho_{i,j;l_1,l_2}^{(1,2)} \rho_{j,i;l_1,l_2}^{(1,2)}) \sqrt{r_{i,i}^{(1)}r_{j,j}^{(1)}r_{i,i}^{(2)}r_{j,j}^{(2)}}.$$
\noindent Henceforth, 
\begin{eqnarray}
  \Theta_{i,j} &=& \theta_{i,j}^{(1)} + \theta_{i,j}^{(2)} - \frac{2}{nq^2} \sum_{l_1,l_2}^q (\rho_{i,i;l_1,l_2}^{(1,2)}\rho_{j,j;l_1,l_2}^{(1,2)} + \rho_{i,j;l_1,l_2}^{(1,2)}\rho_{j,i;l_1,l_2}^{(1,2)})\nonumber\\
               &=& \theta_{i,j}^{(1)} + \theta_{i,j}^{(2)} - \frac{2}{nq}(\rho_{S_{1,2}, i, i}\rho_{S_{1,2}, j, j} + \rho_{S_{1,2}, i, j}\rho_{S_{1,2}, j, i})\cdot \|\Rho_{T_{1,2}}\|_F^2/q.\nonumber
\end{eqnarray}
This completes the proof.

\subsection{Proof of Proposition \ref{lemma:corr_est}} 

Without loss of generality, we assume in our proof that $\omega_{S_t,i,i}=1$ for $i=1,\ldots,p$, $t=1,2$. Let $\teps_{k,i,l}^{(t)} = \epsilon_{k,i,l}^{(t)} - \beps_{k,i,l}^{(t)}$, and the corresponding estimator is defined as $\hat{\epsilon}^{(t)}_{k,i,l}=Y^{(t)}_{k,i,l}-\bar{Y}^{(t)}_{\cdot,i,\cdot}-(\Y^{(t)}_{k,-i,l}-\bar{\Y}^{(t)}_{\cdot,-i,\cdot})^{\T}\hat{\be}_{i}^{(t)}$, with $\bar{\Y}_{\cdot,-i,\cdot}^{(t)} = \frac{1}{nq}\sum_{k=1}^n\sum_{l=1}^q \Y_{k,i,l}^{(t)}$. By  \eqref{eq:inv_reg} in the paper, we have, uniformly in $1\leq i, j \leq p$,
  \begin{eqnarray}
    \heps_{k,i,l}^{(1)}\heps_{k,j,l}^{(2)} &=& \teps_{k,i,l}^{(1)}\teps_{k,j,l}^{(2)} - (\Y_{k,-j,l}^{(2)} - \bar{\Y}_{\cdot,-j,\cdot}^{(2)})(\hbe_j^{(2)} - \be_j^{(2)})\teps_{k,i,l}^{(1)} - (\Y_{k,-i,l}^{(1)} - \bar{\Y}_{\cdot,-i,\cdot}^{(1)})(\hbe_i^{(1)} - \be_i^{(1)})\teps_{k,j,l}^{(2)} \nonumber\\
    && + (\hbe_i^{(1)} - \be_i^{(1)})^T(\Y_{k, -i,l}^{(1)} - \bar{\Y}_{\cdot,-i,\cdot}^{(1)})^T(\Y_{k, -j,l}^{(2)} - \bar{\Y}_{\cdot,-j,\cdot}^{(2)}) (\hbe_j^{(2)} - \be_j^{(2)}).\label{eq:eps_expand_oracle}
  \end{eqnarray}
Denote $\hSigma^{(1,2)}_{i,j} = \frac{1}{nq} \sum_{k = 1}^n \sum_{l = 1}^q (\Y_{k,i,l}^{(1)} - \bar{\Y}_{\cdot,i,\cdot}^{(1)})^T(\Y_{k,j,l}^{(2)} - \bar{\Y}_{\cdot,j,\cdot}^{(2)})$, and $\S_{i,j}^{(1,2)} = \text{tr}(\Rho_{T_{1,2}})/q \cdot \S_{S_{1,2}, i, j}$.  We discuss each term in equation (\ref{eq:eps_expand_oracle}) separately. Note that
    \begin{eqnarray}
      &&|(\hbe_i^{(1)} - \be_i^{(1)})^T\hSigma_{-i, -j}^{(1, 2)}(\hbe_j^{(2)} - \be_j^{(2)})| \nonumber\\
      &\leq& |(\hbe_i^{(1)} - \be_i^{(1)})^T(\hSigma_{-i, -j}^{(1,2)} - \S_{-i, -j}^{(1, 2)})(\hbe_j^{(2)} - \be_j^{(2)})| + |(\hbe_i^{(1)} - \be_i^{(1)})^T \S_{-i,-j}^{(1,2)}(\hbe_j^{(2)} - \be_j^{(2)})|. \label{ineq:term4}
    \end{eqnarray}
For the first term on the right hand side of \eqref{ineq:term4}, we have that, for any $M > 0$, by (A1), there exists $C > 0$ such that
    \begin{equation*}
      \P(\max_{1\leq i < j \leq p} |\hsigma_{i, j}^{(1,2)} - \sigma_{i, j}^{(1,2)}| \geq C \{\log p/(nq)\}^{1/2}) = O(p^{-M}).
    \end{equation*}
Then Assumption (A5) implies that
    \begin{equation}
      \max_{i, j}|(\hbe_i^{(1)} - \be_i^{(1)})^T(\hSigma_{-i, -j}^{(1,2)} - \S_{-i, -j}^{(1,2)})(\hbe_j^{(2)} - \be_j^{(2)})| \leq O_{\rm p}(a_{n1}^2\cdot \{\log p/(nq)\}^{1/2}).\label{ineq:term41}
    \end{equation}
In addition, by the condition that $\sigma_{\max}(\S_{S_{1,2}})|\text{tr}(\Rho_{T_{1,2}})|/q\leq c_1$ in (A1), it follows that
    \begin{equation}
    |(\hbe_i^{(1)} - \be_i^{(1)})^T\S_{-i, -j}^{(1,2)}(\hbe_j^{(2)} - \be_j^{(2)})| =O(\|\hbe_i^{(1)} - \be_i^{(1)}\|_2\cdot \|\hbe_j^{(2)} - \be_j^{(2)}\|_2).\label{ineq:term42}
  \end{equation}
  
By (\ref{ineq:term4}), (\ref{ineq:term41}), (\ref{ineq:term42}), and Assumption (A5), it follows that
  \begin{eqnarray}
   &&\max_{i,j}\Big |\frac{1}{nq}\sum_{k=1}^{n}\sum_{l = 1}^q(\hbe_i^{(1)} - \be_i^{(1)})^T(\Y_{k, -i,l}^{(1)} - \bar{\Y}_{\cdot,-i,\cdot}^{(1)})^T(\Y_{k, -j,l}^{(2)} - \bar{\Y}_{\cdot,-j,\cdot}^{(2)}) (\hbe_j^{(2)} - \be_j^{(2)})\Big|\cr
   &&\quad
    = O_{\rm p}(a_{n2}^2 + a_{n1}^2\cdot \{\log p/(nq)\}^{1/2}).\label{eq:term4}
  \end{eqnarray}
Next, we control the bound of $\teps_{k,i,l}^{(1)}(\Y_{k, -j,l}^{(2)} - \bar{\Y}_{\cdot,-j,\cdot}^{(2)})(\hbe_j^{(2)} - \be_j^{(2)})$. Let $\nu_{i, j^{\prime}} = \Cov\{ \teps_{k,i,l}^{(1)},(Y_{k, j^{\prime},l}^{(2)} - \bar{Y}_{\cdot,j^{\prime},\cdot}^{(2)}) \}$. By (A1), $\underset{i,j^{\prime}}{\max}~ \nu_{i,j^{\prime}}$ is bounded. Then for any $M > 0$, there exists $C > 0$ such that 
$$\P\left( \max_{i, j^{\prime}}| \frac{1}{nq}\sum_{k = 1}^{n}\sum_{l = 1}^q\teps_{k,i,l}^{(1)}(Y_{k, j^{\prime},l}^{(2)} - \bar{Y}_{\cdot,j^{\prime},\cdot}^{(2)}) - \nu_{i,j^{\prime}}| \geq C \{\log p/(nq)\}^{1/2} \right) = O(p^{-M}).$$
Consequently, we have
  \begin{eqnarray}
   && \max_{i,j}\Big |\frac{1}{nq} \sum_{k = 1}^{n}\sum_{l = 1}^q\teps_{k,i,l}^{(1)}(\Y_{k, -j,l}^{(2)} - \bar{\Y}_{\cdot,-j,\cdot}^{(2)})(\hbe_j^{(2)} - \be_j^{(2)})\Big| \nonumber\\
 &\leq& \max_{i,j}\Big |[\frac{1}{nq}\sum_{k = 1}^{n}\sum_{l = 1}^q \teps_{k,i,l}^{(1)}(\Y_{k, -j,l}^{(2)} - \bar{\Y}_{\cdot,-j,\cdot}^{(2)}) - \Bnu_{i, -j}](\hbe_j^{(2)} - \be_j^{(2)})| + |\Bnu_{i, -j}(\hbe_j^{(2)} - \be_j^{(2)})\Big |\nonumber\\
    &=& O_{\rm p}(a_{n1}(\{\log p/(nq)\}^{1/2} + 1)).\label{eq:term2}
  \end{eqnarray}
Combining (\ref{eq:eps_expand_oracle}), (\ref{eq:term4}), (\ref{eq:term2}), and by symmetry, it follows that, uniformly in $1\leq i, j\leq p$, 
  \begin{eqnarray}
    \frac{1}{nq}\sum_{k = 1}^{n}\sum_{l = 1}^q\heps_{k,i,l}^{(1)}\heps_{k,j,l}^{(2)} &=& \frac{1}{nq}\sum_{k = 1}^{n}\sum_{l = 1}^q\teps_{k,i,l}^{(1)}\teps_{k,j,l}^{(2)} + O_{\rm p}(a_{n2}^2 + a_{n1} + \{\log p/(nq)\}^{1/2}\cdot (a_{n1} + a_{n1}^2)).\nonumber
  \end{eqnarray}
  Note that
  \begin{eqnarray}
    \frac{1}{nq}\sum_{k = 1}^{n}\sum_{l = 1}^q\teps_{k,i,l}^{(1)}\teps_{k,j,l}^{(2)} &=& \EE [\frac{1}{nq}\sum_{k = 1}^{n}\sum_{l = 1}^q\teps_{k,i,l}^{(1)}\teps_{k,j,l}^{(2)}] + O_{\rm p} (\{\log p/(nq)\}^{1/2}) \nonumber\\                     &=& \sqrt{r_{i,i}^{(1)}r_{j,j}^{(2)}}\cdot \rho_{S_{1,2}, i, j} \cdot \text{tr}(\Rho_{T_{1,2}})/q + O_{\rm p} (\{\log p/(nq)\}^{1/2}),\nonumber
  \end{eqnarray}
uniformly for $1\leq i, j \leq p$, and by Lemma \ref{lemma:hatrij_oracle} that $\underset{1 \leq i \leq p}{\max} |\hat{r}_{i,i}^{(t)} - r_{i,i}^{(t)}| = O_{\rm p}[\{\log p/(nq)\}^{1/2}]$. Consequently, uniformly in $1 \leq i,j\leq p$,
$$
\hat{\varrho}_{i,j}^{(1,2)} = \frac{1}{nq}\sum_{k = 1}^{n}\sum_{l = 1}^q\heps_{k,i,l}^{(1)}\heps_{k,j,l}^{(2)}/\sqrt{\hat{r}_{i,i}^{(1)}\hat{r}_{j,j}^{(2)}} = \rho_{S_{1,2},i,j}\cdot \text{tr}(\Rho_{T_{1,2}})/q + O_p(a_{n1}).
$$
By Lemma \ref{lemma:hatrij_oracle} again, we have 
$$
\underset{1\leq i < j \leq p}{\max}|nq(\hat{\theta}_{i,j}^{(t)} - \theta_{i,j}^{(t)})| = o_{\rm p}(1/{\log p}).
$$ 
Thus, under (A1), we have, uniformly for $1 \leq i, j\leq p$, 
\begin{eqnarray*}
  nq\hat{\Theta}_{i,j} &=&  nq \left \{\hat{\theta}_{i,j}^{(1)} + \hat{\theta}_{i,j}^{(2)} - \frac{2}{nq} (\hat{\varrho}_{i,i}^{(1,2)}\hat{\varrho}_{j,j}^{(1,2)}+\hat{\varrho}_{i,j}^{(1,2)}\hat{\varrho}_{j,i}^{(1,2)})\cdot \frac{\|\Rho_{T_{1,2}}\|_F^2/q}{[\text{tr}(\Rho_{T_{1,2}}/q)]^2}\right \}\\
                       &=& nq\theta_{i,j}^{(1)} + nq\theta_{i,j}^{(2)} - nq \cdot \frac{2}{nq^2}(\rho_{S_{1,2}, i,i}\rho_{S_{1,2}, j,j} +\rho_{S_{1,2}, i,j}\rho_{S_{1,2}, j,i})\cdot {\|\Rho_{T_{1,2}}\|_F^2} + o_p(1/\log p) \\
                       &=& nq \Theta_{i,j} + o_p(1/\log p).
\end{eqnarray*}
Equivalently, we have $\underset{i,j}{\max}|nq(\hat{\Theta}_{i,j} - \Theta_{i,j})| = o_p(1/\log p)$. This completes the proof.

\subsection{Proof of Proposition \ref{lemma:data}}

Without loss of generality, we assume in this proof that $\omega_{S_t,i,i}=1$ for $i=1,\ldots,p$, $t=1,2$. The estimator of $\teps_{k,i,l}^{(t)}$ is  $\hat{\epsilon}^{(t,d)}_{k,i,l}=Y^{(t,d)}_{k,i,l}-\bar{Y}^{(t,d)}_{\cdot,i,\cdot}-(\Y^{(t,d)}_{k,-i,l}-\bar{\Y}^{(t,d)}_{\cdot,-i,\cdot})^{\T}\hat{\be}_{i}^{(t,d)}$. We have
  \begin{eqnarray*}
\heps_{k,i,l}^{(1,d)}\heps_{k,j,l}^{(2,d)} &=& \teps_{k,i,l}^{(1)}\teps_{k,j,l}^{(2)} - (\Y_{k,-j,l}^{(2)} - \bar{\Y}_{\cdot,-j,\cdot}^{(2)})(\hbe_j^{(2,d)} - \be_j^{(2)})\teps_{k,i,l}^{(1)} - (\Y_{k,-i,l}^{(1)} - \bar{\Y}_{\cdot,-i,\cdot}^{(1)})(\hbe_i^{(1,d)} - \be_i^{(1)})\teps_{k,j,l}^{(2)} \nonumber\\
                                     && + (\hbe_i^{(1,d)} - \be_i^{(1)})^T(\Y_{k, -i,l}^{(1)} - \bar{\Y}_{\cdot,-i,\cdot}^{(1)})^T(\Y_{k, -j,l}^{(2)} - \bar{\Y}_{\cdot,-j,\cdot}^{(2)}) (\hbe_j^{(2,d)} - \be_j^{(2)})\non
                                       && + \bigg[ \{Y_{k,i,l}^{(1,d)} - \bar{Y}_{\cdot,i,\cdot}^{(1,d)} - (\Y_{k,-i,l}^{(1,d)} - \bar{\Y}_{\cdot,-i,\cdot}^{(1,d)})^T\hbe_i^{(1,d)}\}\{Y_{k,j,l}^{(2,d)} - \bar{Y}_{\cdot,j,\cdot}^{(2,d)} - (\Y_{k,-j,l}^{(2,d)} - \bar{\Y}_{\cdot,-j,\cdot}^{(2,d)})^T\hbe_j^{(2,d)}\} \non
                                     && - \{Y_{k,i,l}^{(1)} - \bar{Y}_{\cdot,i,\cdot}^{(1)} - (\Y_{k,-i,l}^{(1)} - \bar{\Y}_{\cdot,-i,\cdot}^{(1)})^T\hbe_i^{(1,d)}\}\{Y_{k,j,l}^{(2)} - \bar{Y}_{\cdot,j,\cdot}^{(2)} - (\Y_{k,-j,l}^{(2)} - \bar{\Y}_{\cdot,-j,\cdot}^{(2)})^T\hbe_j^{(2,d)}\} \bigg].\nonumber                                      
  \end{eqnarray*}
  It has been shown in the proofs of \textit{Theorems 5 and 6} in \citet{xia2017one} that, uniformly for $1 \leq i, j\leq p$, 
  \begin{eqnarray}
  &&\bigg[ \{Y_{k,i,l}^{(1,d)} - \bar{Y}_{\cdot,i,\cdot}^{(1,d)} - (\Y_{k,-i,l}^{(1,d)} - \bar{\Y}_{\cdot,-i,\cdot}^{(1,d)})^T\hbe_i^{(1,d)}\}\{Y_{k,j,l}^{(2,d)} - \bar{Y}_{\cdot,j,\cdot}^{(2,d)} - (\Y_{k,-j,l}^{(2,d)} -n \bar{\Y}_{\cdot,-j,\cdot}^{(2,d)})^T\hbe_j^{(2,d)}\} \non
    && - \{Y_{k,i,l}^{(1)} - \bar{Y}_{\cdot,i,\cdot}^{(1)} - (\Y_{k,-i,l}^{(1)} - \bar{\Y}_{\cdot,-i,\cdot}^{(1)})^T\hbe_i^{(1,d)}\}\{Y_{k,j,l}^{(2)} - \bar{Y}_{\cdot,j,\cdot}^{(2)} - (\Y_{k,-j,l}^{(2)} - \bar{\Y}_{\cdot,-j,\cdot}^{(2)})^T\hbe_j^{(2,d)}\} \bigg] = o_{\rm p}(a_{n2}^2).\nonumber
  \end{eqnarray}
Then following similar arguments as in the proof of Proposition \ref{lemma:corr_est} and by Assumption (A6), we can show that, uniformly for $1 \leq i, j\leq p$, 
  \begin{eqnarray}
    && \frac{1}{nq} \sum_{k = 1}^n \sum_{l = 1}^q \heps_{k,i,l}^{(1,d)} \heps_{k,j,l}^{(2,d)}\nonumber\\
    &=&\frac{1}{nq}\sum_{k = 1}^n\sum_{l= 1}^q\teps_{k,i,l}^{(1)}\teps_{k,j,l}^{(2)} - (\Y_{k,-j,l}^{(2)} - \bar{\Y}_{\cdot,-j,\cdot}^{(2)})(\hbe_j^{(2,d)} - \be_j^{(2)})\teps_{k,i,l}^{(1)} - (\Y_{k,-i,l}^{(1)} - \bar{\Y}_{\cdot,-i,\cdot}^{(1)})(\hbe_i^{(1,d)} - \be_i^{(1)})\teps_{k,j,l}^{(2)} \non
    && + (\hbe_i^{(1,d)} - \be_i^{(1)})^T(\Y_{k, -i,l}^{(1)} - \bar{\Y}_{\cdot,-i,\cdot}^{(1)})^T(\Y_{k, -j,l}^{(2)} - \bar{\Y}_{\cdot,-j,\cdot}^{(2)}) (\hbe_j^{(2,d)} - \be_j^{(2)}) +  o_{\rm p}(a_{n2}^2)\non
    &=& \frac{1}{nq}\sum_{k = 1}^{n}\sum_{l = 1}^q \teps_{k,i,l}^{(1)}\teps_{k,j,l}^{(2)} + O_{\rm p}(a_{n2}^2 + a_{n1} + \{\log p/(nq)\}^{1/2}\cdot (a_{n1} + a_{n1}^2)) \nonumber\\
    &=& \sqrt{r_{i,i}^{(1)}r_{j,j}^{(2)}}\cdot \rho_{S_{1,2}, i,j} \cdot \text{tr}(\Rho_{T_{1,2}})/q  + O_{\rm p}(a_{n2}^2 + a_{n1} + \{\log p/(nq)\}^{1/2}\cdot(a_{n1} + a_{n1}^2)). \nonumber
  \end{eqnarray}
Next, let
  $$\hat{\S}_{T_{1,2}} = \frac{1}{np} \sum_{k = 1}^n \sum_{i = 1}^p\left [\left \{\X_{k, i,\cdot}^{(1)} - \frac{1}{np}\sum_{k =1}^n\sum_{i=1}^p \X_{k, i, \cdot}^{(1)}\right \}^{\T}\left \{\X_{k, i,\cdot}^{(2)} - \frac{1}{np}\sum_{k =1}^n\sum_{i=1}^p \X_{k, i, \cdot }^{(2)}\right\} \right ].$$
Note that $\EE ( \hat{\S}_{T_{1,2}} ) = \S_{T_{1,2}}\cdot \text{tr}(\S_{S_{1,2}})/p$. By the definition of $\hat{\Rho}_{T_{1,2}}^{(d)}$ in \eqref{eq:est_tildeT} of the paper, it follows that 
\begin{eqnarray*}
  &&\left|\text{tr}(\hat{\Rho}_{T_{1,2}}^{(d)})/q - \text{tr}(\Rho_{T_{1,2}})/q \cdot \text{tr}(\S_{S_{1,2}})/p \right| \non
  &\leq& \left \Vert \hat{\Rho}_{T_{1,2},i,j}^{(d)} - \Rho_{T_{1,2}, i,j} \cdot \text{tr}(\S_{S_{1,2}})/p \right \Vert_{\max} \non
  &=&  \left \Vert \hat{\S}_{T_1}^{-1/2} \hat{\S}_{T_{1,2}} \hat{\S}_{T_2}^{-1/2} - \S_{T_1}^{-1/2} [\S_{T_{1,2}}\cdot \text{tr}(\S_{S_{1,2}})/p ] \S_{T_2}^{-1/2} \right \Vert_{\max} \non
  &=&  \Big\| \{\hat{\S}_{T_1}^{-1/2} - \S_{T_1}^{-1/2} + \S_{T_1}^{-1/2}\} \cdot \{\hat{\S}_{T_{1,2}} - \S_{T_{1,2}}\cdot \text{tr}(\S_{S_{1,2}})/p + \S_{T_{1,2}}\cdot \text{tr}(\S_{S_{1,2}})/p\} \cr
    &&\quad\cdot \{\hat{\S}_{T_2}^{-1/2} - \S_{T_2}^{-1/2} + \S_{T_2}^{-1/2}\}  - \S_{T_1}^{-1/2} \{\S_{T_{1,2}}\cdot \text{tr}(\S_{S_{1,2}})/p\} \S_{T_2}^{-1/2} \Big\|_2,
\label{eq:tr_diff1}
  \end{eqnarray*}  
  where $\|\cdot\|_{\max}$ denotes the element-wise maximum norm.
Then Assumption (A7) implies that
\begin{eqnarray*}
  && \Vert \hat{\S}_{T_t}^{-1/2} - c\S_{T_t}^{-1/2} \Vert_{\max} \leq \Vert \hat{\S}_{T_t}^{-1/2} - c\S_{T_t}^{-1/2} \Vert_2 \leq \Vert \hat{\S}_{T_t}^{-1/2} - c\S_{T_t}^{-1/2} \Vert_{F} \leq q\Vert\hat{\S}_{T_t}^{-1/2} - c\S_{T_t}^{-1/2} \Vert_{\max}\non
  &=& o_{\text{p}}([nq\log p \{\log \max (p, q, n)\}^2]^{-1/4}).\label{eq:root_inv_hat_T_rate}
\end{eqnarray*}
Thus we have
\begin{eqnarray*}
  &&\left|\text{tr}(\hat{\Rho}_{T_{1,2}}^{(d)})/q - \text{tr}(\Rho_{T_{1,2}})/q \cdot \text{tr}(\S_{S_{1,2}})/p \right|  \cr
  &&\leq \left \Vert \{\hat{\S}_{T_1}^{-1/2} - \S_{T_1}^{-1/2}\} \cdot \{\hat{\S}_{T_{1,2}} - \S_{T_{1,2}}\cdot \text{tr}(\S_{S_{1,2}})/p\}\cdot \{\hat{\S}_{T_2}^{-1/2} - \S_{T_2}^{-1/2} \}\right \Vert_{2} \non
  &&\quad + \left \Vert \S_{T_1}^{-1/2} \cdot \{\hat{\S}_{T_{1,2}} - \S_{T_{1,2}}\cdot \text{tr}(\S_{S_{1,2}})/p\}\cdot \{\hat{\S}_{T_2}^{-1/2} - \S_{T_2}^{-1/2} \}\right \Vert_{2}\non
  &&\quad + \left \Vert \{\hat{\S}_{T_1}^{-1/2} - \S_{T_1}^{-1/2}\} \cdot [\S_{T_{1,2}}\cdot \text{tr}(\S_{S_{1,2}})/p]\cdot \{\hat{\S}_{T_2}^{-1/2} - \S_{T_2}^{-1/2} \}\right \Vert_{2}\non
  &&\quad + \left \Vert \{\hat{\S}_{T_1}^{-1/2} - \S_{T_1}^{-1/2}\} \cdot \{\hat{\S}_{T_{1,2}} - \S_{T_{1,2}}\cdot \text{tr}(\S_{S_{1,2}})/p\}\cdot \S_{T_2}^{-1/2}\right \Vert_{2}\non
  &&\quad + \left \Vert \{\S_{T_1}^{-1/2} - \S_{T_1}^{-1/2}\} \cdot [\S_{T_{1,2}}\cdot \text{tr}(\S_{S_{1,2}})/p]\cdot \S_{T_2}^{-1/2}\right \Vert_{2} \cr
  &&\quad+ \left \Vert \S_{T_1}^{-1/2} \cdot \{\hat{\S}_{T_{1,2}} - \S_{T_{1,2}}\cdot \text{tr}(\S_{S_{1,2}})/p\}\cdot \S_{T_2}^{-1/2} \right \Vert_{2} \non
  &&\quad + \left \Vert \S_{T_1}^{-1/2} \cdot [\S_{T_{1,2}}\cdot \text{tr}(\S_{S_{1,2}})/p]\cdot \{\hat{\S}_{T_2}^{-1/2} - \S_{T_2}^{-1/2}\}\right \Vert_{2}\cr
  &&\leq o_{\text{p}}(1/\log p) +\left \Vert \S_{T_1}^{-1/2} \cdot \{\hat{\S}_{T_{1,2}} - \S_{T_{1,2}}\cdot \text{tr}(\S_{S_{1,2}})/p\}\cdot \S_{T_2}^{-1/2} \right \Vert_{2}.
  \label{eq:tr_diff2}
  \end{eqnarray*}
Define 
  \[
  \hat \S_{T_{t_1,t_2}} =  \frac{1}{np} \sum_{k = 1}^n \sum_{i = 1}^p\left [\left \{\X_{k, i,\cdot}^{(t_1)} - \frac{1}{np}\sum_{k =1}^n\sum_{i=1}^p \X_{k, i, \cdot}^{(t_1)}\right \}^{\T}\left \{\X_{k, i,\cdot}^{(t_2)} - \frac{1}{np}\sum_{k =1}^n\sum_{i=1}^p \X_{k, i, \cdot }^{(t_2)}\right\} \right ], t_1,t_2=1,2.
  \]
By the assumption that $\{ \X^{(1)}, \X^{(2)} \}$ follows a matrix normal distribution, together with Assumption (A1), we have 
\begin{eqnarray*}\label{l2bound}
&&\left \Vert \hat{\S}_{T_{1,2}} - \S_{T_{1,2}}\cdot \text{tr}(\S_{S_{1,2}})/p \right \Vert_{2} \cr
&&\quad \leq 
\Bigg\|\left(
\begin{array}{cc}
\hat \S_{T_{1,1}} &    \hat \S_{T_{1,2}}\\
\hat \S_{T_{2,1}} & \hat \S_{T_{2,2}}
\end{array}
\right) - \left(
\begin{array}{cc}
 \S_{T_{1}}\cdot \text{tr}(\S_{S_{1}})/p &   \S_{T_{1,2}}\cdot \text{tr}(\S_{S_{1,2}})/p\\
\S_{T_{1,2}}^{\T}\cdot \text{tr}(\S_{S_{1,2}})/p &  \S_{T_{2}}\cdot \text{tr}(\S_{S_{2}})/p
\end{array}
\right)\Bigg\|_2\cr
 &&\quad = O_{\text{p}}(\sqrt{q/np}).
\end{eqnarray*}
This implies that $|\text{tr}(\hat{\Rho}_{T_{1,2}}^{(d)})/q - \text{tr}(\Rho_{T_{1,2}})/q \cdot \text{tr}(\S_{S_{1,2}})/p|=O_{\text{p}}(\sqrt{q/np})$.

By the arguments above, it follows that
\[
\|\hat{\Rho}_{T_{1,2}}^{(d)} - \Rho_{T_{1,2}} \cdot \text{tr}(\S_{S_{1,2}})/p\|_F^2/q \leq \|\hat{\Rho}_{T_{1,2}}^{(d)} - \Rho_{T_{1,2}} \cdot \text{tr}(\S_{S_{1,2}})/p\|_2^2 = O_{\rm p} (q/(np)).
\]
Thus, under Assumption (A3), along with the condition that  $|\text{tr}(\Rho_{T_{1,2}})\text{tr}(\S_{S_{1,2}})|/(pq)\geq c_1^{-1}$, we have
  \begin{eqnarray}
    \frac{\|\hat{\Rho}_{T_{1,2}}^{(d)}\|_F^2/q}{\bigg \{\text{tr}(\hat{\Rho}_{T_{1,2}}^{(d)})/q\bigg \}^2} =\frac{\|{\Rho}_{T_{1,2}}\|_F^2/q}{\bigg \{\text{tr}({\Rho}_{T_{1,2}})/q\bigg \}^2} + o_{\rm p}(1/\log p).\nonumber
  \end{eqnarray}
 
Finally, by Proposition \ref{lemma:corr_est}, Lemma \ref{lemma:hatrij_data}, and Assumption (A1), we have
\begin{eqnarray}
  nq\hat{\Theta}_{i,j}^{(d)} &=&  nq \left \{\hat{\theta}_{i,j}^{(1,d)} + \hat{\theta}_{i,j}^{(2,d)} - \frac{2}{nq} (\hat{\varrho}_{i,i}^{(1,2,d)}\hat{\varrho}_{j,j}^{(1,2,d)}+\hat{\varrho}_{i,j}^{(1,2,d)}\hat{\varrho}_{j,i}^{(1,2,d)})\cdot \frac{\|\hat{\Rho}_{T_{1,2}}^{(d)}\|_F^2/q}{[\text{tr}(\hat{\Rho}_{T_{1,2}}^{(d)}/q)]^2}\right \}\nonumber\\
                       &=& nq\theta_{i,j}^{(1)} + nq\theta_{i,j}^{(2)} - nq \cdot \frac{2}{nq^2}(\rho_{S_{1,2}, i,i}\rho_{S_{1,2}, j,j} +\rho_{S_{1,2}, i,j}\rho_{S_{1,2}, j,i})\cdot \|\Rho_{T_{1,2}}\|_F^2 + o_p(1/\log p)\nonumber\\
                       &=& nq \Theta_{i,j} + o_p(1/\log p),\nonumber
\end{eqnarray}
uniformly in $1 \leq i, j\leq p$. Equivalently, we have $\underset{i,j}{\max}|nq(\hat{\Theta}_{i,j}^{(d)} - \Theta_{i,j})| = o_p(1/\log p)$. This completes the proof.

\subsection{Proof of Theorem \ref{thm:fdr_control}}

This theorem can be proved by utilizing the same arguments as in the proofs of Theorems 1 and 2 in \citet{xia2018two}, with some  modifications that we discuss below. Without loss of generality, we assume in this proof that $\omega_{S_t, i, i} = 1$ for $t = 1, 2$, $i = 1, \dots, p$. 
  
When $\S_{T_t}$ and $\S_{T_{1,2}}$ are known, for $t = 1, 2$, let $V_{i,j} = (U_{i,j}^{(2)} - U_{i,j}^{(1)})/\Theta_{i,j}^{1/2}$. By Proposition \ref{lemma:corr_est}, we have   
  \begin{equation*}
  \underset{1\leq i < j \leq p}{\max}|nq(\hat{\Theta}_{i,j} - \Theta_{i,j})| = o_{\rm p}(1/{\log p}).
  \end{equation*}
  Also note that for $(i,j)\in \mathcal{H}_0 \setminus A_{\tau}$, we have $|\omega_{S_t,i,j}|=o\{(\log p)^{-1}\}$. Then by Lemma \ref{lemma:hatrij_oracle} and Assumptions (A1) and (A2),
  it is easy to see that, for $(i,j)\in \mathcal{H}_0 \setminus A_{\tau}$ we have, 
  \[
    \max_{(i,j)\in \mathcal{H}_0 \setminus A_{\tau}}\|W_{i,j}|-|V_{i,j}\|=o_{\rm p}\{(\log p)^{-1/2}\}.
  \]
Let $\omega_{S_1,i,j} = \omega_{S_2, i,j} = \omega_{i,j}$ under $H_{0,i,j}$. For $(i,j)\in A_{\tau}$, by Lemma \ref{lemma:hatrij_oracle}, we have
  \[
    W_{i,j}=V_{i,j}+b_{i,j}+o_{\rm p}(\log p^{-1/2}),
  \]
where $b_{i,j}={2\{\omega_{i,j}(\hat{\sigma}^{(1)}_{i,i,\epsilon}-\hat{\sigma}^{(2)}_{i,i,\epsilon})+\omega_{i,j}(\hat{\sigma}^{(1)}_{j,j,\epsilon}-\hat{\sigma}^{(2)}_{j,j,\epsilon})\}}/{{\hat{\Theta}_{i,j}}^{1/2}}$.
Note that, uniformly in $1 \leq i, j\leq p$,
\begin{eqnarray*}
  |b_{i,j}|&\leq& 2 C^{1\over2} \left[ \frac{\sqrt{nq} \cdot |\tilde{\sigma}^{(1)}_{i,i,\epsilon}-\tilde{\sigma}^{(2)}_{i,i,\epsilon}|}{\sqrt{\Var\bigg ((\epsilon^{(1)}_{k,i,l})^2 - (\epsilon^{(2)}_{k,i,l})^2\bigg )}} 
  + \frac{\sqrt{nq}\cdot|\tilde{\sigma}^{(1)}_{j,j,\epsilon}-\tilde{\sigma}^{(2)}_{j,j,\epsilon}|}{\sqrt{\Var \bigg ((\epsilon^{(1)}_{k,j,l})^2 - (\epsilon^{(2)}_{k,j,l})^2\bigg )}} \right]+o_{\rm p}\{(\log p)^{-{1/2}}\},
\end{eqnarray*}
where $\tilde{\sigma}^{(t)}_{i,i,\epsilon}=(nq)^{-1}\sum_{k=1}^{n}\sum_{l=1}^q(\epsilon^{(t)}_{k,i,l})^2$, and $C$ is a bounded constant depending only on $nq\Theta_{i,j}$. Thus, we have
\begin{eqnarray*}
\P\left( \max_{(i,j)\in A_{\tau}}W_{i,j}^2\geq 4\log p-\log \log p+h \right) 
  \leq \text{Card}(A_{\tau})\{\P(V_{i,j}^2\geq \log p/8)+\P(b_{i,j}^2\geq 2\log p)\}=o(1),
\end{eqnarray*}
where the last equality is a direct result of Lemma \ref{lemma:gaussian_ldb}. By the proof of Theorem 1 in \citet{xia2018two}, this conclusion indicates that the set $A_{\tau}$ is negligible, and it suffices to focus on $\mathcal{H}_0 \setminus A_{\tau}$.

Next, we re-define $V_m$'s and $\hat{V}_m$'s used in Theorem 1 in \citet{xia2018two} as follows. We arrange the indices  $\{(i,j): (i,j)\in \mathcal{H}_0\setminus A_{\tau}\}$ in any ordering and set them as $\{(i_{m},j_{m}): m=1,\dots,s\}$ with with $s=$Card$(\mathcal{H}_0 \setminus A_{\tau})$. Let $\Theta_{m}=\Var(\epsilon^{(1)}_{k,i_m,l}\epsilon^{(1)}_{k,j_m,l} - \epsilon^{(2)}_{k,i_m,l}\epsilon^{(2)}_{k,j_m,l})$,  and define $Z_{k,m,l} =(\epsilon^{(2)}_{k,i_{m},l}\epsilon_{k,j_{m},l}^{(2)} -\epsilon^{(1)}_{k,i_{m},l}\epsilon_{k,j_{m},l}^{(1)})-\EE(\epsilon^{(2)}_{k,i_{m},l}\epsilon_{k,j_{m},l}^{(2)}-\epsilon^{(1)}_{k,i_{m},l}\epsilon_{k,j_{m},l}^{(1)})$ for $1\leq k\leq n$ and $1\leq l\leq q$. Define
\[
V_{m} =  \frac{1}{\sqrt{nq\Theta_{m}}}\sum_{k=1}^{n}\sum_{l=1}^qZ_{k,m,l}, \quad \text{ and } \quad 
\hat{V}_{m} = \frac{1}{\sqrt{nq\Theta_{m}}} \sum_{k=1}^{n}\sum_{l=1}^q\hat{Z}_{k,m,l},
\]
where $\hat{Z}_{k,m,l}=Z_{k,m,l}I(|Z_{k,m,l}|\leq \tau_{n})-\EE \{Z_{k,m,l}I(|Z_{k,m,l}|\leq \tau_{n})\}$, and $\tau_{n}=32\log (p+nq)$.

Note that $\max_{(i,j)\in \mathcal{H}_0\setminus A_{\tau}}V^{2}_{i,j}=\max_{1\leq m\leq s}V_{m}^{2}$, and that
   \begin{eqnarray*}
     &&\max_{1\leq m\leq s}(nq)^{-{1/ 2}}\sum_{k=1}^{n_1+n_2}\sum_{l=1}^q\EE[|Z_{k,m,l}|I\{|Z_{k,m,l}|\geq 32\log (p+nq)\}]\cr
     &&\quad\leq C(nq)^{1/ 2}\max_{1\leq k\leq n}\max_{1\leq l\leq q}\max_{1\leq m\leq s}\EE[|Z_{k,m,l}|I\{|Z_{k,m,l}|\geq 32\log (p+nq)\}]\cr
     &&\quad\leq C(nq)^{1/ 2}(p+nq)^{-4}\max_{1\leq k\leq n}\max_{1\leq l\leq q}\max_{1\leq m\leq s}\EE[|Z_{k,m,l}|\exp\{|Z_{k,m,l}|/8\}]\cr
     &&\quad\leq C(nq)^{1/ 2}(p+nq)^{-4}.
   \end{eqnarray*}
This leads to 
   \[
     \P\left\{ \max_{1\leq m\leq s}|V_{m}-\hat{V}_{m}|\geq (\log p)^{-1} \right\} \leq \P \left( \max_{1\leq m\leq s}\max_{1\leq k\leq n}\max_{1\leq l\leq q}|Z_{k,m,l}|\geq \tau_{n} \right) =O(p^{-1}),
   \]
which indicates that $V_m$ behaves almost the same as $\hat{V}_m$, and so does the corresponding $W_{i,j}$. Finally, following the proof of Theorem 1 in \citet{xia2018two} with the redefined $V_m$ and $\hat{V}_m$, we obtain the desired result.    

When $\S_{T_t}$ and $\S_{T_{1,2}}$ are unknown, following the proof of the known $\S_{T_t}$ and $\S_{T_{1,2}}$ scenario, together with Lemma \ref{lemma:hatrij_data} and Proposition \ref{lemma:data}, completes the proof.

\section{Parameter tuning}
\label{appendix:tune_Lasso}

For the estimation of $\be_{i}^{(t)}$ in \eqref{eq:inv_reg} in Section \ref{sec:method:test_stat}, we propose to use Lasso. We adopt a similar tuning procedure for Lasso as developed in \citet{xia2017one}. The idea is to make the number of false rejections $\sum_{(i,j) \in \mathcal{H}_0} I(|W_{i,j}|\geq h)$ and the estimator $\{2 - 2\Phi(h)\}(p^2 - p)/2$ close. Specifically, 
\begin{enumerate}[{Step} 1]
\item Let $\lambda_{n, i}^{(t)} = b/20 \sqrt{\hat{\Sigma}_{S_t, i,i} \log p/(nq)}$, for $b = 1, \ldots ,40$. For each $b$, calculate $\hat\be_{i}^{(t)}$, 
  $i = 1, \ldots , p$, and construct the corresponding standardized statistics $W_{i,j}^{(b)}$, $t = 1, 2$.
\item Choose $\hat{b}$ as the minimizer of
  $$\sum_{s = 1}^{10} \left (\frac{\sum_{i, j\in \mathcal{H}} I (|W_{i,j}^{(b)}| \geq \Phi^{-1}[1 - s\{1 - \Phi(\sqrt{\log p})\}/10])}{s\{1 - \Phi (\sqrt{\log p})\}/10 \cdot p(p-1)} - 1\right)^2.$$
\item The tuning parameters $\lambda_{n,i}^{(t)}$ are then set as, $\lambda_{n,i}^{(t)} = \hat{b}/20\{\hat{\Sigma}_{S_t, i,i} \log p/(nq)\}^{1/2}$, $t = 1,2$.
\end{enumerate}

\section{Estimation of $\Theta_{i,j}$}
\label{appendix:theta_mse}

To complement the simulations in Section \ref{sec:normal}, we report the mean squared error (MSE) of the estimator $\hat{\Theta}^{(d)}$ averaged over all pairs of $(i, j)$ with $i < j$. That is, for each pair of $(i, j)$, $1\leq i<j\leq p$, we first compute the MSE of $\hat{\Theta}_{i,j}^{(d)}$ based on $n$ replications. We then calculate $\sum_{i,j}\text{MSE}(\hat{\Theta}_{i,j}^{(d)})/\sum_{i,j} \Theta_{i,j}^2$, i.e., the average MSE over the average true $\Theta_{i,j}^2$. Table \ref{tbl_appendix:theta_mse} reports the results. It is seen that the estimated $\hat{\Theta}^{(d)}$ with variance correction achieves a smaller estimation error than that without variance correction. This observation holds true across all scenarios, especially when the correlations are strong before and after the stimulus.

\begin{sidewaystable}[hptb]\small\addtolength{\tabcolsep}{-4pt}
\begin{center}
    \caption{The MSE of the estimator $\hat{\Theta}^{(d)}$ averaged over all pairs of $(i, j)$ with $i < j$.}
    \label{tbl_appendix:theta_mse}
    \begin{lrbox}{\tablebox}
    \begin{tabular}{c@{\hspace{1em}} |c@{\hspace{1em}}  |r@{\hspace{1em}} r@{\hspace{1em}}|r@{\hspace{1em}} r @{\hspace{1em}}|r@{\hspace{1em}}r@{\hspace{1em}} |r@{\hspace{1em}} r@{\hspace{1em}}|r@{\hspace{1em}} r @{\hspace{1em}}|r@{\hspace{1em}}r@{\hspace{1em}} }
    \hline
    \multicolumn{2}{c|}{temporal structure}&\multicolumn{6}{c}{moving average}&\multicolumn{6}{|c}{autoregressive}\\[0pt]
    \hline
    \multicolumn{2}{c|}{spatial structure} &\multicolumn{2}{c}{banded}&\multicolumn{2}{|c}{hub}&\multicolumn{2}{|c}{small}&\multicolumn{2}{|c}{banded}&\multicolumn{2}{|c}{hub}&\multicolumn{2}{|c}{small}\\  [0pt]
    \hline
    \multicolumn{2}{c|}{correction}&\multicolumn{1}{c|}{\cmark}&\multicolumn{1}{c|}{\xmark}&\multicolumn{1}{c|}{\cmark}&\multicolumn{1}{c|}{\xmark}&\multicolumn{1}{c|}{\cmark}&\multicolumn{1}{c|}{\xmark}&\multicolumn{1}{c|}{\cmark}&\multicolumn{1}{c|}{\xmark}&\multicolumn{1}{c|}{\cmark}&\multicolumn{1}{c|}{\xmark}&\multicolumn{1}{c|}{\cmark}&\multicolumn{1}{c|}{\xmark}\\[0pt]
    \hline\hline\hline
    $(p, q)$  & $\gamma$ &\multicolumn{12}{c}{Setting I}\\[0pt]
    \hline
    \multirow{5}{*}{$(200, 50)$}  
    &$0.$ &0.0& 0.0  & 0.0  & 0.0 & 0.0 & 0.0 & 0.0 & 0.0 & 0.0 & 0.0  & 0.0  & 0.0 \\[0pt]
    &$0.2$ &0.0& 0.2  & 0.0  & 0.2 & 0.0 & 0.2 & 0.0 & 0.2 & 0.0 & 0.2  & 0.0  & 0.2 \\[0pt]
    &$0.4$ &0.1& 3.5  & 0.1  & 3.6 & 0.1 & 3.5 & 0.1 & 3.5 & 0.1 & 3.6  & 0.1  & 3.5 \\[0pt]
    &$0.6$ &1.5& 31.0  & 0.6  & 31.2 & 1.4 & 31.0 & 1.5 & 30.9 & 0.6 & 31.2  & 1.4  & 31.0 \\[0pt]
    \hline
    \multirow{5}{*}{$(200, 200)$} 
    &$0.$ &0.0& 0.0  & 0.0  & 0.0 & 0.0 & 0.0 & 0.0 & 0.0 & 0.0 & 0.0  & 0.0  & 0.0 \\[0pt]
    &$0.2$ &0.0& 0.2  & 0.0  & 0.2 & 0.0 & 0.2 & 0.1 & 0.2 & 0.0 & 0.2  & 0.0  & 0.2 \\[0pt]
    &$0.4$ &0.1& 3.5  & 0.1  & 3.6 & 0.1 & 3.5 & 0.1 & 3.5 & 0.1 & 3.6  & 0.1  & 3.5 \\[0pt]
    &$0.6$ &1.5& 31.0  & 0.7  & 31.2 & 1.5 & 31.0 & 1.5 & 31.0 & 0.7 & 31.2  & 1.5  & 31.1 \\[0pt]
    \hline
    \multirow{5}{*}{$(800, 200)$} 
    &$0.$ &0.0& 0.0  & 0.0  & 0.0 & 0.0 & 0.0 & 0.0 & 0.0 & 0.0 & 0.0  & 0.0  & 0.0 \\[0pt]
    &$0.2$ &0.0& 0.2  & 0.0  & 0.2 & 0.0 & 0.2 & 0.0 & 0.2 & 0.0 & 0.2  & 0.0  & 0.2 \\[0pt]
    &$0.4$ &0.0& 3.6  & 0.3  & 3.6 & 0.0 & 3.6 & 0.0 & 3.6 & 0.3 & 3.6  & 0.0  & 3.6 \\[0pt]
    &$0.6$ &2.4& 31.5  & 2.9  & 31.5 & 1.9 & 31.5 & 2.4 & 31.5 & 2.9 & 31.5  & 1.8  & 31.5 \\[0pt]    
    \hline
    & &\multicolumn{12}{c}{Setting II}\\[0pt]
    \hline
    \multirow{5}{*}{$(200, 50)$}  
    &$0.$ &0.0& 0.0  & 0.0  & 0.0 & 0.0 & 0.0 & 0.0 & 0.0 & 0.0 & 0.0  & 0.0  & 0.0 \\[0pt]
    &$0.2$ &0.0& 0.8  & 0.0  & 0.3 & 0.0 & 0.5 & 0.0 & 0.8 & 0.0 & 0.3  & 0.0  & 0.5 \\[0pt]
    &$0.4$ &3.7& 12.4  & 0.8  & 4.5 & 3.8 & 7.9 & 3.8 & 12.4 & 0.8 & 4.5  & 3.8  & 7.9 \\[0pt]
    &$0.6$ &28.4& 44.5  & 10.6  & 19.6 & 23.2 & 31.2 & 28.4 & 44.5 & 10.6 & 19.6  & 23.2  & 31.2 \\[0pt]
    \hline
    \multirow{5}{*}{$(200, 200)$}
    &$0.$ &0.0& 0.0  & 0.0  & 0.0 & 0.0 & 0.0 & 0.0 & 0.0 & 0.0 & 0.0  & 0.0  & 0.0 \\[0pt]
    &$0.2$ &0.1& 0.8  & 0.0  & 0.3 & 0.0 & 0.5 & 0.1 & 0.8 & 0.0 & 0.3  & 0.0  & 0.5 \\[0pt]
    &$0.4$ &3.0& 12.4  & 0.6  & 4.5 & 2.9 & 7.9 & 3.0 & 12.4 & 0.6 & 4.5  & 3.0  & 7.9 \\[0pt]
    &$0.6$ &26.5& 44.5  & 10.0  & 19.6 & 21.3 & 31.2 & 26.6 & 44.5 & 10.0 & 19.6  & 21.3  & 31.2 \\[0pt]    
    \hline
    \multirow{5}{*}{$(800, 200)$}
    &$0.$ &0.0& 0.0  & 0.0  & 0.0 & 0.0 & 0.0 & 0.0 & 0.0 & 0.0 & 0.0  & 0.0  & 0.0 \\[0pt]
    &$0.2$ &0.0& 0.9  & 0.0  & 0.2 & 0.0 & 0.5 & 0.0 & 0.9 & 0.0 & 0.2  & 0.0  & 0.5 \\[0pt]
    &$0.4$ &4.2& 12.9  & 1.1  & 3.3 & 3.8 & 7.6 & 4.2 & 12.9 & 1.1 & 3.3  & 3.8  & 7.6 \\[0pt]
    &$0.6$ &29.8& 45.6  & 9.9  & 14.8 & 22.8 & 30.4 & 29.8 & 45.6 & 9.9 & 14.8  & 22.8  & 30.4 \\[0pt]    
    \hline
    \end{tabular}
    \end{lrbox}
    \scalebox{0.8}{\usebox{\tablebox}}
\end{center}
\end{sidewaystable}


\baselineskip=16pt
\bibliographystyle{biorefs}
\bibliography{ref-testpair}

\end{document}